\documentclass[twocolumn,showpacs,preprintnumbers,amsmath,amssymb,prb,floatfix]{revtex4}

\usepackage{graphicx}
\usepackage{dcolumn}
\usepackage{bm}
\usepackage{longtable}

\begin{document}

\title{Nonequilibrium perturbation theory of the spinless
Falicov-Kimball model}

\author{V.~Turkowski} \email{turkowskiv@missouri.edu}
\altaffiliation{New address: Department of Physics and Astronomy,
University of Missouri-Columbia, Columbia, MO 65211}

\author{J.K.~Freericks}\email{jkf@physics.georgetown.edu}

 \affiliation{Department of Physics, Georgetown University,
Washington, D.C. 20057 USA}

\date{\today}

\begin{abstract}
We perform a perturbative analysis for the nonequilibrium Green functions
of the spinless Falicov-Kimball model
in the presence of an arbitrary external time-dependent
but spatially uniform electric field. The conduction electron self-energy
is found from a strictly truncated second-order perturbative expansion in
the local electron-electron repulsion $U$.  We examine the
current at half-filling, and compare to both the semiclassical
Boltzmann equation and exact numerical solutions for the contour-ordered
Green functions from a transient-response formalism (in infinite dimensions)
on the Kadanoff-Baym-Keldysh contour.  We find a strictly
truncated perturbation theory in the two-time formalism
cannot reach the long-time limit of the steady state;
instead it illustrates pathological behavior for times larger than
approximately $2/U$.
\end{abstract}



\maketitle

\section{Introduction}

The theoretical description of the physical properties of strongly
correlated electrons in an external time-dependent electric field is
an important problem in condensed matter physics. One reason is the
potential for using strongly correlated materials in electronics,
due to their tunability. These materials have unusual and
interesting properties, which can be modified by slightly varying
physical parameters, like temperature, pressure, or carrier
concentration. For example, the conductance of nanoelectronic
devices can be controlled by tuning the carrier concentration and
applying an electric field\cite{Ahn}. Moreover, because of the small
size of nanoelectronic devices, an electric potential of the order
of ${\rm 1~V}$ produces electric fields on the order of ${\rm
10^{5}-10^{6}~V/cm}$ or higher. Therefore, it is important to
understand the response of such systems to a strong electric field.
Recent progress in nanotechnology has resulted in many experimental
results on strongly correlated electron nanostructures in
time-dependent electric fields which must also be understood. One
such system is the quantum dot, which can be described by a Kondo
impurity attached to two leads, through two tunnel
junctions\cite{Doyon,andrei_bethe}. Other examples of important
experiments on strongly correlated systems in time-dependent
electric fields include an unusually strong change in the optical
transmission in the transition metal oxide\cite{Ogasawara} ${\rm
Sr_{2}CuO_{3}}$. and the dielectric breakdown of a Mott insulator
that occurs in the quasi-one-dimensional cuprates ${\rm
Sr_{2}CuO_{3}}$, and ${\rm SrCuO_{2}}$, when a strong electric field
is applied \cite{Taguchi}. Generally speaking, since most electron
devices have nonlinear current-voltage characteristics, it is
important to understand how electron correlations will modify this
effect.

Due to the complexity of these problems, there is a dearth of
theoretical results on different models available. Similar to
equilibrium, the simplest cases to analyze are the one-dimensional
case and the approximation of infinite dimensions, where dynamical
mean-field theory (DMFT) can be applied. The zero dimensional
problem (quantum dot) in the case of a Kondo impurity coupled to two
leads, or sources of electrons, has also been studied by different
groups. In particular, the case of a single-impurity Anderson model
was examined in Refs.~\onlinecite{Nordlander1,Nordlander2,Coleman}
and a Bethe ansatz approach with open boundary conditions has now
been developed for similar problems\cite{andrei_bethe}. An attempt
to describe optical transmission experiments\cite{Ogasawara} in
${\rm SrCuO_{3}}$ was performed in the framework of the
one-dimensional two-band Hubbard model, where the problem on a
twelve site periodic ring was solved exactly. In
Ref.~\onlinecite{SchmidtMonien}, the spectral properties of the
Hubbard model coupled to a periodically time varying chemical
potential were studied to second order in the Coulomb repulsion with
iterated perturbation theory. The possibility of the breakdown of a
Mott insulator in the 1D Hubbard model was discussed in
Ref.~\onlinecite{Oka} by numerically solving the time-dependent
Schroedinger equation for the many-body wave function. The
electrical and the spectral properties of the infinite-dimensional
spinless Falicov-Kimball model in an external
homogeneous electric field have also been examined \cite{Nashville,Turkowski1,%
dmft_fk}.

The spinless Falicov-Kimball model
is one of the simplest models for strongly correlated electrons
that acquires the main features of a correlated electron system.
It consists of two kinds of electrons: conducting $c$-electrons
and localized $f$-electrons. These two different electrons
interact through an on-site Coulomb repulsion.
The model can be regarded as a simplified version of the
Hubbard model, where the spin-up {($c$)} electrons move
in the background of the frozen spin-down {($f$)} electrons.
The Falicov-Kimball model was originally introduced as a
model to describe valence-change and metal-insulator
transitions\cite{FalicovKimball},
and later was interpreted as a model for crystallization or
binary alloy formation\cite{KennedyLieb}.
Much progress has been made on solving this model with DMFT in equilibrium,
where essentially all properties of the conduction electrons in equilibrium
are known (for a review, see Ref.~\onlinecite{Freericks}).

We have already studied the nonequilibrium response
of the conduction electrons in the Falicov-Kimball model
to a uniform electric field turned on at a particular moment of time
with an exact numerical DMFT formalism\cite{Nashville,Turkowski1,dmft_fk}.
We solved a system of generalized nonequilibrium DMFT
equations for the electron Green function and self-energy
and found interesting effects caused by the electric field.
In particular, we saw how the conduction-electron density of states (DOS)
evolved from a Wannier-Stark ladder of delta functions  at the Bloch
frequencies to broadened
peaks with maxima located away from the Bloch frequencies.
Another interesting phenomenon we found was that the
Bloch oscillations of the electric current can
survive out to long times in the interacting case, when the amplitude
of the
electric field is large. These results were determined by numerically
solving a system of the equations for the Green function
and self-energy defined on a complex time contour
(see Fig.~\ref{fig: keldysh}), according
to the Kadanoff-Baym-Keldysh nonequilibrium Green function
formalism\cite{kadanoff_baym,keldysh}.
The numerical results were obtained for a finite time interval
because computer resources restrict the contour to be a finite length.
Therefore, the steady state could not be rigorously reached.

Our original motivation for this work, was to find a simple perturbation
theory that could reach the steady state, at least for weak electron-electron
interactions.  Surprisingly, we found that a transient-based perturbation
theory breaks down once the time becomes larger than $2/U$, as described
in detail below.  Nevertheless, even though the perturbation theory cannot give
a definite answer about the steady state of the system,
it does allow us to study the physical properties of the system
in the transient regime, and for small $U$, we can extend the work farther
than can be done with the exact numerical techniques. Therefore,
it might help us understand the evolution of the system
toward the steady state in the case when the electron correlations
are weak.  In this contribution,
we calculate the response of the Falicov-Kimball
model to an external electric field by calculating the strictly truncated
electron self-energy to second order in $U$.
We focus on the time-dependence of the electric current and the density
of states
in the case of a constant electric field turned on at time $t=0$.

The paper is organized as follows: We define the problem in Section
II and introduce the nonequilibrium dynamical mean-field theory
formalism to solve it in Section III. In Section IV, the expressions
for the nonequilibrium Green functions are derived. These functions
are used to study the time-dependence of the electric current in
Section V and the electron density of states in Section VI.
Conclusions and discussions are presented in Section VII.

\section{Falicov-Kimball model}
The Hamiltonian for the spinless Falicov-Kimball model has the following
form:
\begin{equation}
{\cal H}=-\sum_{\langle ij\rangle}t_{ij}c_{i}^{\dagger}c^{}_{j}
-\mu\sum_{i}c_{i}^{\dagger}c^{}_{i}
-\mu_{f}\sum_{i}f_{i}^{\dagger}f^{}_{i}
+U\sum_{i}f_{i}^{\dagger}f^{}_{i}c_{i}^{\dagger}c^{}_{i}.
\label{H}
\end{equation}
The Falicov Kimball model has two kinds of spinless
electrons: conduction $c$-electrons and localized $f$-electrons.
The nearest-neighbor hopping matrix element $t_{ij}=t^*/2\sqrt{d}$
is written in a form appropriate
for the infinite-dimensional limit where $d$ is the spatial dimension of the
hypercubic lattice and $t^*$ is a rescaled hopping matrix
element\cite{metzner_vollhardt}; note that our formal results hold
in any dimension, but all numerical calculations are performed in the
infinite-dimensional limit. The summation is over nearest neighbors $i$
and $j$.  The on-site Coulomb repulsion
between the two kinds of the electrons is equal to $U$.
The symbols $\mu$ and $\mu_{f}$ denote the chemical potentials
of the conduction and the localized electrons, respectively.
For our numerical work, we will consider the case of half-filling, where
the particle densities of the conduction and localized electrons
are each equal to $0.5$ and $\mu =\mu_{f}=U/2$.

We study the response of the system
to an external electromagnetic field ${\bf E}({\bf r},t)$.
In general, an external electromagnetic field is expressed
by a scalar potential $\varphi ({\bf r},t)$ and by a
vector potential ${\bf A}({\bf r},t)$ in the following way:
\begin{equation}
{\bf E}({\bf r}, t)=-{\bf \nabla}\varphi ({\bf r}, t)
-\frac{1}{c}\frac{\partial {\bf A}({\bf r}, t)}
{\partial t}.
\label{Electricfield}
\end{equation}
For simplicity, we assume that the electric field is spatially uniform.
In this case, it is convenient
to choose the temporal or Hamiltonian gauge for the electric field:
$\varphi ({\bf r}, t)=0$.
Then, the electric field is described
by a spatially homogeneous vector potential ${\bf A}(t)$,
and the Hamiltonian
maintains its translational invariance, so it
has a simple form in the momentum
representation [see Eq.~(\ref{Hk}) below].
This choice of the gauge allows one to avoid additional
complications in the calculations caused by
the inhomogeneous scalar potential.
The assumption of a spatially homogeneous electric field
is equivalent to neglecting all magnetic field effects
produced by the vector potential
[since ${\bf B}({\bf r},t)={\bf \nabla}\times {\bf A}({\bf r}, t)$].
This approximation is valid when the vector potential
is smooth enough, that the magnetic field produced by ${\bf A}({\bf r}, t)$
can be neglected.  When the electric field is described
by a vector potential ${\bf A}({\bf r}, t)$ only,
the electric field is coupled to the conduction electrons
by means of the Peierls substitution
for the hopping matrix\cite{Jauho}:
\begin{eqnarray}
t_{ij}&\rightarrow& t_{ij}\exp\left[
-\frac{ie}{\hbar c}\int_{{\bf R}_{i}}^{{\bf R}_{j}}{\bf A}({\bf r},
t)d{\bf r}
\right]\nonumber\\
&=&t_{ij}\exp\left[
\frac{ie}{\hbar c}
{\bf A}(t)\cdot ({\bf R}_{i}-{\bf R}_{j}).
\right] ,
\label{Peierls1}
\end{eqnarray}
where the second equality follows for a spatially uniform electric field.

The Peierls substituted Hamiltonian in Eq.~(\ref{H}) assumes a
simple form in momentum space:
\begin{eqnarray}
H(A)
&=&\sum_{{\bf k}}\left[\epsilon \left({\bf k}-\frac{e{\bf A}(t)}{\hbar
c}\right)
-\mu\right] c_{{\bf k}}^{\dagger}c_{{\bf k}}^{}
-\mu_f\sum_{\bf k}f^\dagger_{\bf k}f^{}_{\bf k}\nonumber\\
&+&
U\sum_{{\bf p},{\bf k},{\bf q}}
f_{{\bf p}+{\bf q}}^{\dagger}c_{{\bf k}-{\bf q}}^{\dagger}
c_{{\bf k}}^{}f_{{\bf p}}^{};
\label{Hk}
\end{eqnarray}
in Eq.~(\ref{Hk}), the creation and annihilation operators are expressed in a
momentum-space basis.
The free electron energy spectra in Eq.~(\ref{Hk}) is
\begin{equation}
\epsilon\left({\bf k}-\frac{e{\bf A}(t)}{\hbar c}\right)
=-2t\sum_{l=1}^{d}\cos\left[a\left( k^{l}-\frac{eA^{l}(t)}{\hbar c}
\right)\right] .
\label{Ek}
\end{equation}
We shall study the case when the electric field (and the vector potential)
lies along the elementary cell diagonal
\cite{Turkowski}:
\begin{equation}
{\bf A}(t)=A(t)(1,1,...,1).
\label{A}
\end{equation}
This form is convenient for calculations,
since in this case the spectrum of the non-interacting electrons
coupled to the electric field becomes
\begin{eqnarray}
\epsilon \left({\bf k}-\frac{e{\bf A}(t)}{\hbar c}\right)
&=&\cos\left(\frac{eaA(t)}{\hbar c}\right)\varepsilon ({\bf k})\nonumber\\
&+&\sin\left(\frac{eaA(t)}{\hbar c} \right){\bar \varepsilon} ({\bf k}),
\label{energy}
\end{eqnarray}
which depends on only two energy functions:
\begin{equation}
\varepsilon ({\bf k})=-2t\sum_{l}\cos (ak^{l})
\label{eps}
\end{equation}
and
\begin{equation}
{\bar \varepsilon} ({\bf k})=-2t\sum_{l}\sin (ak^{l}).
\label{bareps}
\end{equation}
The former is the noninteracting electron bandstructure
(in zero electric field), and the latter
is the sum of the components of the noninteracting electron
velocity multiplied by $-1$.

The diagonal electric field is especially convenient in
the limit of infinite dimensions $d\rightarrow\infty$,
when the electron self-energies are local in space,
or momentum-independent.  Then the
joint density of states for the two energy functions in Eqs.~(\ref{eps})
and (\ref{bareps}) can be directly found.
It has a double Gaussian form for the infinite-dimensional hypercubic lattice
\cite{Schmidt}:
\begin{eqnarray}
\rho_{2}(\epsilon , {\bar \varepsilon})
=\frac{1}{\pi t^{*2}a^{d}}\exp\left[
-\frac{\varepsilon^{2}}{t^{*2}}-\frac{{\bar \epsilon}^{2}}{t^{*2}}
\right] .
\label{rho2}
\end{eqnarray}
Below we use the units, where $a=\hbar =c=t^{*}=1$.

\section{Nonequilibrium formalism}

\begin{figure}[htb]
\centering{
\includegraphics[width=8.0cm]{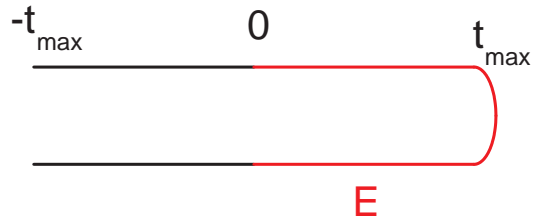}
}
\caption{
Kadanoff-Baym-Keldysh
time contour for the two-time contour-ordered Green function
in nonequilibrium. The red line corresponds
to the region of time where the electric field is nonzero ($t>0$).
}
\label{fig: keldysh}
\end{figure}

The nonequilibrium properties of the system in Eq.~(\ref{Hk}) can be
studied by
calculating the contour-ordered Green function:
\begin{eqnarray}
&~&G_{{\bf k}}^{c}(t_{1},t_{2})=
-i
\langle {\rm {\hat T}}_{c}c_{{\bf k}H}^{}(t_{1})
c_{{\bf k}H}^{\dagger}(t_{2})\rangle\label{GT}\\
&=&\frac{-i{\rm Tr}\left\{e^{-\beta H(-t_{\rm max})}
{\rm {\hat T}_{c}}\exp [-i\int_{c}dt H_I(t)]
c_{{\bf k}I}^{}(t_{1})c_{{\bf k}I}^{\dagger}(t_{2})\right\}}
{{\rm Tr}e^{-\beta H(-t_{\rm max})}},
\nonumber
\end{eqnarray}
where the time integration is performed
along ``the interaction'' time contour depicted in Fig.~\ref{fig: keldysh}
(see, for example, Ref.~\onlinecite{Rammer}) and the time ordering
${\rm {\hat T}_{c}}$ is with respect to times along the contour. The angle
brackets indicate that one is taking the trace over states weighted by
the equilibrium density matrix $\exp[-\beta H(-t_{\rm max})]/Z$ with
$Z={\rm Tr}\exp[-\beta H(-t_{\rm max})]$; this is done because the system
is initially prepared in equilibrium prior to the field being turned on.

The indices $H$ and $I$
on the right hand side of Eq.~(\ref{GT}) stand for
the Heisenberg and Interaction representations
for the electron operators.
$H(-t_{\rm max})$ corresponds to the Hamiltonian
in the case of zero electric field.
The expression in Eq.~(\ref{GT}) is a formal generalization
of the equilibrium formulas to the nonequilibrium case.
Therefore,
all the machinery of equilibrium quantum many-body theory,
including Wick's theorem and the different relations
between the Green functions, the self-energies and the vertex functions
can be directly applied to the nonequilibrium case as long as we replace
time-ordered objects by the contour-ordered objects along the contour
of Fig.~\ref{fig: keldysh}
(see, for example, Ref.~\onlinecite{Rammer} for an appropriate discussion).

In this work, we use the two-branch Keldysh
contour, which is appropriate when we directly solve the problem on the
lattice (rather than using DMFT techniques, where the contour is
necessarily more complicated).
This contour consists of two horizontal branches:
one runs to the right and the other to the left.  Such a time contour is needed
to be able to calculate both time-ordered and so-called lesser or greater
Green functions, because it allows us to use time ordering along the
whole contour  to determine those objects.  Because of the time ordering,
we can use Wick's theorem to evaluate different correlation functions for
the perturbative expansion.  In general, there are four different Green
functions one defines by taking the time coordinates along the upper $+$
or lower $-$ branch.  These are the time-ordered, lesser, greater, and
anti-time-ordered Green functions.  They satisfy
\begin{eqnarray}
G^T_{\bf k}(t_1,t_2)&=&G^c_{\bf k}(t_1^+,t_2^+)
\nonumber\\
&=&-i\langle{\rm  T}c_{{\bf k}H}^{}(t_1)c^\dagger_{{\bf k}H}(t_2)
\rangle,\label{g_t}\\
G^<_{\bf k}(t_1,t_2)&=&G^c_{\bf k}(t_1^+,t_2^-)
\nonumber\\
&=&i\langle c^\dagger_{{\bf k}H}(t_2)c_{{\bf k}H}^{}(t_1)
\rangle,\label{g_lesser}\\
G^>_{\bf k}(t_1,t_2)&=&G^c_{\bf k}(t_1^-,t_2^+)
\nonumber\\
&=&-i\langle c_{{\bf k}H}^{}(t_1)c^\dagger_{{\bf k}H}(t_2)
\rangle,\label{g_greater}\\
G^{\bar T}_{\bf k}(t_1,t_2)&=&G^c_{\bf k}(t_1^-,t_2^-)
\nonumber\\
&=&-i\langle{\rm \bar T}c_{{\bf k}H}^{}(t_1)c^\dagger_{{\bf k}H}(t_2)
\rangle,
\label{eq: g_anti_t}
\end{eqnarray}
where the $+$ and $-$ signs indicate whether the real times $t_1$ and $t_2$
lie on the upper (positive time direction)
or lower (negative time direction) branch of the time contour, the symbol T
is ordinary time ordering and the symbol ${\rm \bar T}$ is anti time ordering.
The more familiar retarded and advanced Green functions satisfy
\begin{eqnarray}
G^R_{\bf k}(t_1,t_2)&=&G^T_{\bf k}(t_1,t_2)-G^<_{\bf k}(t_1,t_2)\nonumber\\
&=&-G^{\bar T}_{\bf k}(t_1,t_2)+G^>_{\bf k}(t_1,t_2),
\label{eq: g_ret}\\
G^A_{\bf k}(t_1,t_2)&=&G^T_{\bf k}(t_1,t_2)-G^>_{\bf k}(t_1,t_2)\nonumber\\
&=&-G^{\bar T}_{\bf k}(t_1,t_2)+G^<_{\bf k}(t_1,t_2),
\label{eq: g_adv}
\end{eqnarray}
and the so-called Keldysh Green function is
\begin{eqnarray}
G^K_{\bf k}(t_1,t_2)&=&G^<_{\bf k}(t_1,t_2)+G^>_{\bf k}(t_1,t_2)\nonumber\\
&=&G^T_{\bf k}(t_1,t_2)+G^{\bar T}_{\bf k}(t_1,t_2).
\label{eq: g_keldysh}
\end{eqnarray}
It is sometimes convenient to introduce a two-component
Fermi-field operator
\begin{equation}
{\tilde c}_{{\bf k}}(t) =
\left(
\begin{array}{c}
c_{{\bf k}}(t^{+})\\
c_{{\bf k}}(t^{-})
\end{array}
\right) , \ \ \ \ {\tilde c}_{{\bf k}}^{\dagger}(t) = \left( c_{{\bf
k}}^{\dagger}(t^{+}),c_{{\bf k}}^{\dagger}(t^{-}) \right)
\label{cpm}
\end{equation}
with the components also defined
on the upper and lower time branches of the time contour.
In this case, the following matrix Green function can be introduced:
\begin{equation}
{\hat G}_{{\bf k}}^{c}(t_{1},t_{2}) =
\left(
\begin{array}{cc}
G_{{\bf k}}(t_{1}^{+},t_{2}^{+})\\
G_{{\bf k}}(t_{1}^{-},t_{2}^{+})
\end{array}
\begin{array}{cc}
G_{{\bf k}}(t_{1}^{+},t_{2}^{-})\\
G_{{\bf k}}(t_{1}^{-},t_{2}^{-}).
\end{array}
\right)
\label{Gpm}
\end{equation}
In order to transform the matrix in Eq.~(\ref{Gpm}) to a form more
convenient for computation and often used to study nonequilibrium
problems, we make the combined Keldysh and
Larkin-Ovchinnikov (LO) transformation \cite{Larkin,Rammer}:
\begin{equation}
{\hat G}_{{\bf k}}^{T}(t_{1},t_{2}) \rightarrow {\hat L}{\hat
\tau}_{z}{\hat G}_{{\bf k}}^{T}(t_{1},t_{2}) {\hat L}^{\dagger},
\label{LarkinOvchinnikov}
\end{equation}
where ${\hat L}=(1/\sqrt{2})({\hat \tau}_{0}-i{\hat \tau}_{y})$,
and ${\hat \tau}_{i}, i=0,x,y,z$, are the Pauli matrices.
Under such a transformation, the Green function in Eq.~(\ref{Gpm})
acquires the following form:
\begin{equation}
{\hat G}_{{\bf k}}^{c}(t_{1},t_{2})=
\left(
\begin{array}{c}
G_{{\bf k}}^{R}(t_{1},t_{2})\\
0
\end{array}
\begin{array}{c}
G_{{\bf k}}^{K}(t_{1},t_{2})\\
G_{{\bf k}}^{A}(t_{1},t_{2})
\end{array}
\right) ,
\label{Gbasis}
\end{equation}
where the retarded, advanced and the Keldysh  Green functions are
also expressed as
\begin{equation}
G_{\bf k}^{R}(t_{1},t_{2})
=-i\theta (t_{1}-t_{2})\left\langle
\left\{ c_{{\bf k}H}^{}(t_{1}),c_{{\bf
k}H}^{\dagger}(t_{2})\right\}\right\rangle ,
\label{GR}
\end{equation}
\begin{equation}
G_{\bf k}^{A}(t_{1},t_{2})
=i\theta (t_{2}-t_{1})\left\langle
\left\{ c_{{\bf k}H}^{}(t_{1}),c_{{\bf
k}H}^{\dagger}(t_{2})\right\}\right\rangle ,
\label{GA}
\end{equation}
\begin{equation}
G_{\bf k}^{K}(t_{1},t_{2})
=-i\left\langle \left[
c_{{\bf k}H}^{}(t_{1}),c_{{\bf k}H}^{\dagger}(t_{2})
\right]
\right\rangle .
\label{GK}
\end{equation}
The square (curly) brackets  correspond to the commutator
(anti-commutator).  In this representation, the time arguments
of Green functions (\ref{GR})-(\ref{GK}) are real times.

In the nonequilibrium case, all the information about the properties
of a system is contained in two basic electron Green functions. The
most often used ones are the retarded function in Eq.~(\ref{GR}) and
the lesser function in Eq.~(\ref{g_lesser}).

The retarded Green function contains information about the spectra
of the system, and the lesser Green function describes the
occupation of these states. These two functions form an independent
Green function basis, and any other Green function can be expressed
by means of these two functions; in the equilibrium case, only one
Green function is independent (since the states are distributed
according to the Fermi-Dirac distribution). In particular, it is
easy to find a useful relation, which follows from the definitions
in Eqs.~(\ref{GR}), (\ref{GA}), (\ref{GK}) and (\ref{g_lesser}):
\begin{equation}
G_{\bf k}^{<}(t_{1},t_{2})
=
\frac{1}{2}\left(
G_{\bf k}^{K}(t_{1},t_{2})
-G_{\bf k}^{R}(t_{1},t_{2})
+G_{\bf k}^{A}(t_{1},t_{2})
\right) .
\label{GlGK}
\end{equation}

The self-energy representation, which corresponds to Eq.~(\ref{Gbasis})
has the same form:
\begin{equation}
{\hat \Sigma}_{{\bf k}}^{c}(t_{1},t_{2})=
\left(
\begin{array}{c}
\Sigma^{R}_{{\bf k}}(t_{1},t_{2})\\
0
\end{array}
\begin{array}{c}
\Sigma^{K}_{{\bf k}}(t_{1},t_{2})\\
\Sigma^{A}_{{\bf k}}(t_{1},t_{2})
\end{array}
\right) .
\label{Sigmabasis}
\end{equation}
In this representation one can also apply the standard Wick rules
to evaluate the different operator averages of the perturbative expansion.
In fact, it is possible to show that the nonequilibrium and equilibrium
field theories are formally equivalent if one introduces
the time ordering along the contour instead
of the usual ``equilibrium'' time ordering\cite{Langreth}.

The Dyson equation which connects the contour-ordered
Green function in Eq.~(\ref{Gbasis}) and self-energy, remains valid:
\begin{equation}
{\hat G}_{{\bf k}}^{c}(t_{1},t_{2})
={\hat G}_{{\bf k}}^{0c}(t_{1},t_{2})
+\left[
{\hat G}_{{\bf k}}^{0c}{\hat \Sigma}_{{\bf k}}^{c}{\hat G}_{{\bf
k}}^{c}
\right](t_{1},t_{2}),
\label{Dyson}
\end{equation}
where ${\hat G}_{{\bf k}}^{0c}(t_{1},t_{2})$ is the nonequilibrium
Green function in the case of no interactions, and the product of the
three operators in the last term is a shorthand notation for an
implicit matrix multiplication of the continuous matrix operators over their
respective internal time coordinates (evaluated along the contour).
Its components can be found analytically. For the conduction electrons, we have
\begin{eqnarray}
G_{{\bf k}}^{R0}(t_{1},t_{2})&=&-i\theta (t_{1}-t_{2})\exp [i\mu^{(0)}(t_1-t_2)]
\nonumber\\
&\times&\exp \left[
-i\int_{t_{2}}^{t_{1}}d{\bar t}\epsilon \left({\bf k}-e{\bf A}({\bar
t}) \right) \right] , \label{GR0}\\
G_{{\bf k}}^{A0}(t_{1},t_{2})&=&i\theta (t_{2}-t_{1})\exp [i\mu^{(0)}(t_1-t_2)]
\nonumber\\
&\times&\exp \left[
-i\int_{t_{2}}^{t_{1}}d{\bar t}\epsilon \left({\bf k}-e{\bf A}({\bar
t}) \right) \right] , \label{GA0}\\
G_{{\bf k}}^{K0}(t_{1},t_{2})&=&i\{2f[\epsilon ({\bf k})-\mu^{(0)}]-\}1
\exp [i\mu^{(0)}(t_1-t_2)]
\nonumber\\
&\times& \exp
\left[ -i\int_{t_{2}}^{t_{1}}d{\bar t}\epsilon \left({\bf k}-e{\bf
A}({\bar t}) \right) \right] , \label{GK0}
\end{eqnarray}
where $f[\epsilon ({\bf k})-\mu^{(0)}]=1/\{1+\exp[\beta\{\epsilon ({\bf k})
-\mu^{(0)}\}]\}$
is the Fermi-Dirac distribution function for free electrons. The symbol
$\mu^{(0)}$ is the noninteracting chemical potential.
While the $f$-electron Green functions are:
\begin{equation}
F_{{\bf k}}^{R0}(t_{1},t_{2})
=-i\theta (t_{1}-t_{2})e^{i\mu_f^{(0)}(t_1-t_2)},
\label{FR0}
\end{equation}
\begin{equation}
F_{{\bf k}}^{A0}(t_{1},t_{2})
=i\theta (t_{2}-t_{1})e^{i\mu_f^{(0)}(t_1-t_2)},
\label{FA0}
\end{equation}
\begin{equation}
F_{{\bf k}}^{K0}(t_{1},t_{2})=i(2n_f-1)e^{i\mu_f^{(0)}(t_1-t_2)}.
\label{FK0}
\end{equation}
Here, $\mu_f^{(0)}$ is the noninteracting chemical potential for the
$f$-electrons and $n_f$ is the concentration of $f$-electrons.  As expected,
the free $f$-electron Green functions are momentum-independent. Note that at
half-filling, we have $\mu^{(0)}=\mu_f^{(0)}=0$ and $n_f=1/2$, so that
the noninteracting Green functions simplify, and in particular
$F_{{\bf k}}^{K0}=0$.

The Dyson equation in Eq.~(\ref{Dyson}) is equivalent to the following
system of three equations:
\begin{equation}
G_{\bf k}^{R}(t_{1},t_{2})=G_{\bf k}^{R0}(t_{1},t_{2})
+
[G_{\bf k}^{R0}\Sigma_{{\bf k}}^{R}G_{\bf k}^{R}](t_{1},t_{2}) ,
\label{DysonGR}
\end{equation}
\begin{equation}
G_{\bf k}^{A}(t_{1},t_{2})=G_{\bf k}^{A0}(t_{1},t_{2}) +[G_{\bf
k}^{A0}\Sigma_{{\bf k}}^{A}G_{\bf k}^{A}](t_{1},t_{2}) ,
\label{DysonGA}
\end{equation}
\begin{eqnarray}
G_{\bf k}^{K}(t_{1},t_{2})&=&
[1+G_{\bf k}^{R}\Sigma_{{\bf k}}^{R}]G_{\bf k}^{K0}[1+\Sigma_{\bf
k}^{A}G_{\bf k}^{A}]
(t_{1},t_{2})\nonumber\\
&+&[G_{\bf k}^{R}\Sigma_{{\bf k}}^{K}G_{\bf k}^{A}](t_{1},t_{2}).
\label{DysonGK}
\end{eqnarray}
This is derived by simply writing down the Dyson equation
for every matrix component. The 21 component is trivial,
since every 21 matrix component is equal to zero.
We omit the internal time integrals (over the real-time axis) in
Eqs.~(\ref{DysonGR})-(\ref{DysonGK}) for sake of brevity.

In general, it is difficult to solve this system of equations,
especially in the nonequilibrium case when electron interactions are present.
However, for the Falicov-Kimball model the electron self-energy
is momentum-independent through second order in the interaction $U$,
which simplifies the analysis.
In the next Section, we shall present the perturbative solution
of Eq.~(\ref{Dyson}), or equivalently Eqs.~(\ref{DysonGR})-(\ref{DysonGK}),
for the Green function.

\section{Perturbation theory}

We will investigate a non-self-consistent perturbative expansion that
is strictly truncated to second order in $U$.  We perform the
expansion directly for the lattice Hamiltonian, which is worked out (for the
equilibrium case) in the
Appendix.  As mentioned above, the first and second-order self-energies
are local because the $f$-electron Green function is local in
all dimensions.  The expression for the nonequilibrium
self-energy can be obtained from the corresponding expression for
the equilibrium time-ordered self-energy by making the Langreth
substitution\cite{Langreth}
for the corresponding Green functions, which tells us to replace
all integrals over the real time axis by integrals over the contour, and to
replace all time-ordered objects by contour-ordered objects.  The equilibrium
time-ordered self-energy has the following expression when strictly
truncated to second order in $U$ (see Appendix A)
\begin{eqnarray}
\Sigma_{lm}(t_{1},t_{2}) &=&\delta_{lm}[\delta
(t_{1}-t_{2})(Un_{f}-\mu+\mu^{(0)})\nonumber\\
&+&U^{2}G_{ll}^{0}(t_{1},t_{2})F_{ll}^0(t_{2},t_{1})
F_{ll}^{0}(t_{1},t_{2})]
\label{Sigmaequilibrium}
\end{eqnarray}
where $l$ and $m$ denote lattice sites and we have suppressed the time-ordered
superscript on all quantities. Using the fact that the product of the
$f$-electron Green functions is $n_f(1-n_f)$, yields the final
expression for the second-order equilibrium self-energy
\begin{equation}
\Sigma_{lm}^{(2)}(t_{1},t_{2})=
\delta_{lm}U^2n_f(1-n_f)G_{ll}^{0}(t_{1},t_{2}).
\label{eq: self_energy_2_final}
\end{equation}
Note that this expression cannot immediately be used to find the equilibrium
retarded self-energy.  In order to find that, we can determine the time-ordered
self-energy on the imaginary time axis, Fourier transform to
Matsubara frequencies and then perform an analytic continuation to the real
axis to find the retarded self-energy.  An alternate method, which we will
adopt here, is to find the equilibrium lesser self-energy from the nonequilibrium
formalism (evaluated in the equilibrium limit).  Then the equilibrium retarded
self-energy follows by taking the difference of the equilibrium time-ordered and
lesser self-energies.

In the nonequilibrium case, we use the Langreth rule\cite{Langreth}
to replace the integrals over real time
by integrals over the contour.  Then by examining time arguments
on each of the two branches, and following the same strategy as in
the appendix, we find
\begin{eqnarray}
\Sigma^{T(2)}(t_1,t_2)&=&U^2G^{T0}(t_1,t_2)F^{T0}(t_1,t_2)F^{T0}(t_2,t_1)
\label{eq: sigma_t}\\
\Sigma^{<(2)}(t_1,t_2)&=&U^2G^{<0}(t_1,t_2)F^{<0}(t_1,t_2)F^{>0}(t_2,t_1)
\label{eq: sigma_<}\\
\Sigma^{>(2)}(t_1,t_2)&=&U^2G^{>0}(t_1,t_2)F^{>0}(t_1,t_2)F^{<0}(t_2,t_1)
\label{eq: sigma_>}\\
\Sigma^{\bar T(2)}(t_1,t_2)&=&U^2G^{\bar T0}(t_1,t_2)F^{\bar T0}(t_1,t_2)
F^{\bar T0}(t_2,t_1).
\label{eq: sigma_anti_t}
\end{eqnarray}
Each pair of products of $f$-electron Green functions is equal to
$n_f(1-n_f)$, so after performing the Keldysh and the LO
transformations, we get
\begin{eqnarray}
\left(
\begin{array}{c}
\Sigma^{R(2)}\\
0
\end{array}
\begin{array}{c}
\Sigma^{K(2)}\\
\Sigma^{A(2)}
\end{array}
\right) (t_{1},t_{2}) &=&U^{2}n_f(1-n_f)\label{SigmaPT2}\\
&\times&\left(
\begin{array}{c}
G^{R0}\\
0
\end{array}
\begin{array}{c}
G^{K0}\\
G^{A0}
\end{array}
\right)(t_{1},t_{2}), \nonumber
\end{eqnarray}
where all the Green functions are local.  For instance,
\begin{equation}
G^{R0}(t_{1},t_{2})= \frac{1}{N}\sum_{\bf k}G^{R0}_{{\bf
k}}(t_{1},t_{2}). \label{G0loc}
\end{equation}
This approach also allows us to determine the equilibrium retarded
self-energy by simply
evaluating the results in the absence of an electric field.  This is an example
of how one uses the nonequilibrium technique to complete
an equilibrium analytic continuation.

Now we will focus on the half-filled case [where $n_f(1-n_f)=1/4$,
$\mu^{(0)}=\mu_f^{(0)}=0$, and the first-order self-energy vanishes].
The formal expressions for the Green functions can be found from the
appropriate Dyson equations
[Eqs.~(\ref{DysonGR})-(\ref{DysonGK})].  Here, we write down the results
only for the half-filled case:
\begin{equation}
G_{\bf k}^{R(2)}(t_{1},t_{2})=G_{\bf k}^{R0}(t_{1},t_{2}) + [G_{\bf
k}^{R0}\Sigma^{R(2)}G_{\bf k}^{R0}](t_{1},t_{2}) , \label{DysonGR2}
\end{equation}
\begin{equation}
G_{\bf k}^{A(2)}(t_{1},t_{2})=G_{\bf k}^{A0}(t_{1},t_{2}) +[G_{\bf
k}^{A0}\Sigma^{A(2)}G_{\bf k}^{A0}](t_{1},t_{2}) , \label{DysonGA2}
\end{equation}
\begin{eqnarray}
G_{\bf k}^{K(2)}(t_{1},t_{2})&=& G_{\bf k}^{K0}(t_{1},t_{2}) +[G_{\bf
k}^{R0}\Sigma^{R(2)}G_{\bf k}^{K0} \label{DysonGK2}\\
&+&G_{\bf k}^{K0}\Sigma^{A(2)}G_{{\bf
k}}^{A0} +G_{\bf k}^{R0}\Sigma^{K(2)}G_{\bf k}^{A0}](t_{1},t_{2}),
\nonumber
\end{eqnarray}
where the second-order self-energies $\Sigma^{(2)}$ and the free Green
functions $G^{0}$ are given by the half-filled case in Eqs.~(\ref{SigmaPT2}) and
(\ref{GR0})-(\ref{GK0}), respectively. (In the general case, we need to
also include the first-order self-energy and its iterated second-order
contribution.)

The free $c$-electron local Green functions (in the presence of a
spatially uniform electric field), which are necessary to calculate
the self-energy (\ref{SigmaPT2}), can be found by summing
Eqs.~(\ref{GR0})-(\ref{GK0})  over momentum
[as in Eq.~(\ref{G0loc})]\cite{Turkowski}:
\begin{eqnarray}
G^{R0}(t_{1},t_{2})&=&-i\theta (t_{1}-t_{2})
e^{i\mu^{(0)}(t_1-t_2)}\nonumber\\
&\times&\exp \left\{ -\frac{1}{4}
\left[ \left( \int_{t_{2}}^{t_{1}}d{\bar t} \cos \left( e{\bf
A}({\bar t})\right) \right)^{2} \right . \right .\nonumber\\
&+&\left . \left .\left( \int_{t_{2}}^{t_{1}}d{\bar
t} \sin \left( e{\bf A}({\bar t})\right) \right)^{2} \right]
\right\} , \label{GR0loc}
\end{eqnarray}
\begin{eqnarray}
G^{A0}(t_{1},t_{2})&=&i\theta (t_{2}-t_{1})e^{i\mu^{(0)}(t_1-t_2)}\nonumber\\
&\times&\exp \left\{ -\frac{1}{4}
\left[ \left( \int_{t_{2}}^{t_{1}}d{\bar t} \cos \left( e{\bf
A}({\bar t})\right) \right)^{2} \right . \right .\nonumber\\
&+&\left . \left .\left( \int_{t_{2}}^{t_{1}}d{\bar
t} \sin \left( e{\bf A}({\bar t})\right) \right)^{2} \right]
\right\} , \label{GA0loc}
\end{eqnarray}
\begin{eqnarray}
G^{K0}(t_{1},t_{2}) &=&ie^{i\mu^{(0)}(t_1-t_2)}\nonumber\\
&\times&\exp \left\{ -\frac{1}{4} \left(
\int_{t_{2}}^{t_{1}}d{\bar t} \sin \left( e{\bf A}({\bar t})\right)
\right)^{2} \right\}
\nonumber \\
&~&\times\int d\epsilon \rho (\epsilon ) [2f (\epsilon-\mu^{(0)} )-1] \nonumber\\
&~&\times\exp \left\{-i\epsilon \int_{t_{2}}^{t_{1}}d{\bar t} \cos
\left( e{\bf A}({\bar t})\right) \right\}, \label{GK0loc}
\end{eqnarray}
where
\begin{equation}
\rho (\epsilon ) =\frac{1}{\sqrt{\pi}} e^ {-\varepsilon^{2}}
\label{rho}
\end{equation}
is the free electron density of states on the hypercubic lattice in
infinite dimensions (the generalization to finite dimensions is
obvious). At half-filling we have $\mu^{(0)}=0$ in
Eqs.~(\ref{GR0loc})-(\ref{GK0loc}).

Thus, we have obtained the formal expressions for the second-order
nonequilibrium retarded, advanced and Keldysh Green functions
(\ref{DysonGR})-(\ref{DysonGK}), with the free Green functions
defined by Eqs.~(\ref{GR0})-(\ref{GK0}),
(\ref{GR0loc})-(\ref{GK0loc}) and the self-energies in Eq.~(\ref{SigmaPT2}).
In the following Section, we shall use these
solutions to study the time-dependence of the electric current in
the case when an external time-dependent electric field [Eq.~(\ref{A})]
is present.

Before proceeding with the nonequilibrium solutions we would like to
study some of the equilibrium properties of the retarded self-energy
$\Sigma$. We check
the perturbative result for the self-energy in Eq.~(\ref{SigmaPT2})
versus the exact DMFT results in equilibrium
and estimate the values of $U$ above which the second order
equilibrium perturbation theory fails. It is natural to suppose that
the second order nonequilibrium perturbation theory results are also
inaccurate when $U$ is that large.

\begin{figure}[h]
\centering{
\includegraphics[width=8.0cm,clip=]{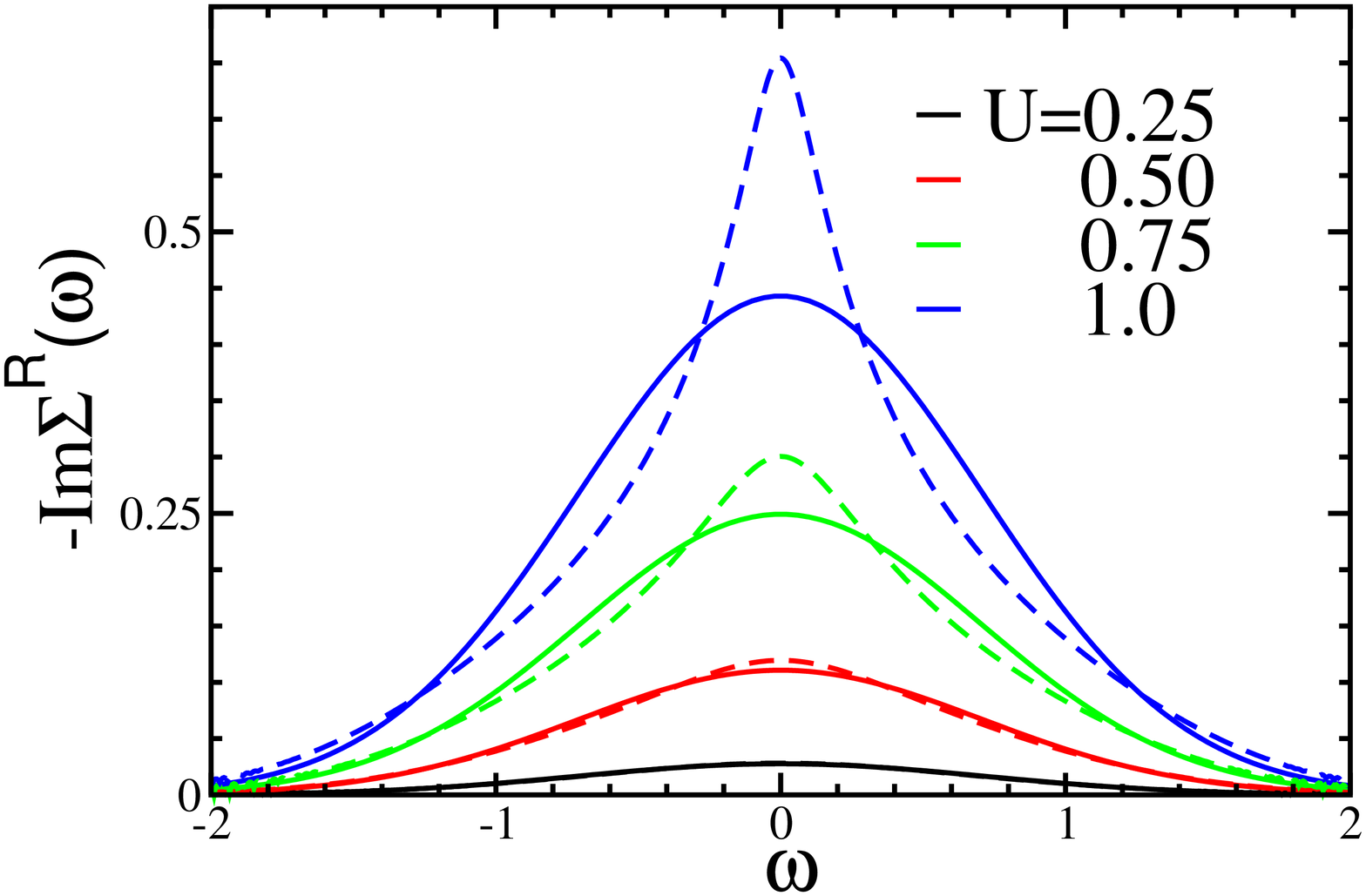}}
\caption{Frequency dependence of the imaginary part of the
second order retarded self-energy for different values of the
Coulomb repulsion $U$ (solid lines). The dashed lines are the
corresponding exact dynamical mean-field theory results. }
\label{fig: self_energy_equilib}
\end{figure}

Using the results in Eq.~(\ref{SigmaPT2}), we find the expression for the
second order self-energy has the following form in frequency
space (at half-filling):
\begin{eqnarray}
\Sigma^{R} (\omega ) &=&\frac{U^{2}}{4}G^{R0}(\omega),\nonumber\\
&=&\frac{U^{2}}{4} \left[{\rm \hat P} \int
d\varepsilon\rho (\epsilon ) \frac{1}{\omega -\varepsilon} -i\pi
\rho (\omega)\right] ,
\label{Sigmaequilibrium2}
\end{eqnarray}
where the symbol ${\rm \hat P}$ denotes the principle value of the
integral, and $\rho (\epsilon )$ is defined in Eq.~(\ref{rho}). In
order to estimate the values of $U$, for which the second order
perturbation theory gives good results in equilibrium, we compare
the imaginary part of the perturbative retarded self-energy
\begin{equation}
{\rm Im}\Sigma^{R}(\omega) =-\frac{\sqrt{\pi}U^{2}}{4} e^{-\omega^{2}}
\label{ImSigmaequilibrium}
\end{equation}
with the exact self-energy calculated by the DMFT approach. The
frequency dependence of $-{\rm Im}\Sigma^{R}$ for different values
of $U$ is presented in Fig.~\ref{fig: self_energy_equilib}. As
follows from this figure, the second order perturbation theory gives
reasonable results up to $U\sim 0.5$.  Hence, we expect that the
result in Eq.~(\ref{SigmaPT2}) will also be accurate when $U<0.5$
for the nonequilibrium case.

\section{The electric current}

\subsection{Semiclassical Boltzmann equation approximation for the current}

Before analyzing the time dependence of the current in the quantum
case, we present the approximation for the current that follows from
a semiclassical Boltzmann equation approach. The Boltzmann equation
for the nonequilibrium quasiparticle distribution function $f^{\rm
non}({\bf k},t)=-iG_{{\bf k}}^{<}(t,t)$ becomes
\begin{equation}
\frac{\partial f^{\rm non}({\bf k},t)}{\partial t} +e{\bf
E}(t)\cdot\nabla_{\bf k}f^{\rm non}({\bf k},t)
=-\frac{1}{\tau}[f^{\rm non}({\bf k},t)-f({\bf k})],
\label{Boltzmann}
\end{equation}
in the case of a spatially uniform time-dependent electric field.
Here, $\tau$ is the scattering time, which typically is proportional
to $1/U^{2}$ in the weak correlated limit of the Falicov-Kimball
model (because it is inversely proportional to the imaginary part of
the self-energy at zero frequency) and the r.h.s. of the equation is
the collision term. Note that in the quantum case, the quantum
Boltzmann equation is more complicated because the quantum
excitations depend on two independent times, not one.  Hence, it isn't obvious
that we should use the quantum relaxation time in the semiclassical Boltzmann
equation; indeed, we often find that by fitting to a somewhat longer
relaxation time, we can improve the fit to the Boltzmann equation
rather dramatically, but we present the results using a fixed relaxation
time here. Of course, our numerical
nonequilibrium results presented below are equivalent to exactly solving the
full quantum Boltzmann equation, but our method of solution is
different.

\begin{figure}[h]
\centering{
\includegraphics[width=8.0cm,clip=]{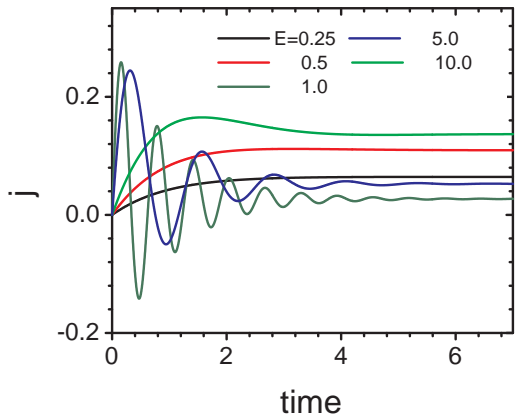}}
\caption{Boltzmann equation solution for the time-dependence of the
electric current for different values of the electric field in the
limit of infinite dimensions at $\tau=1, \beta =10$. Here and in
Figs.~\ref{fig:4}--\ref{fig:7}, we set the prefactor $e/\sqrt{d}$ in
the results for the electric current [see Eqs.~(\ref{jBoltzmannsol})
and (\ref{j3})] to be equal to $1$.} \label{fig: boltzmann}
\end{figure}

The boundary condition for the semiclassical Boltzmann equation is:
\begin{eqnarray}
f^{\rm non}({\bf k},t=0)&=&f({\bf k})\nonumber\\
&=&\frac{1}{\exp[\beta (\varepsilon({\bf k}) -\mu)]+1} .
\label{Boltzmannbc}
\end{eqnarray}
We will be considering only the case of half-filling, where we can
set the chemical potential to zero. The solution of
Eqs.~(\ref{Boltzmann}) and (\ref{Boltzmannbc}) when the electric
field is pointing in the diagonal direction ${\bf
E}(t)=E(1,1,...,1)\theta (t)$ has the following form:
\begin{equation}
f^{\rm non}({\bf k},t) =\frac{1}{\tau}\int_{-\infty}^{t}d{\bar t}
e^{-(t-{\bar t})/\tau}f\left ({\bf k}+e\int_{t}^{{\bar t}}dt^\prime
{\bf E}(t^\prime)\right ). \label{Boltzmannsol}
\end{equation}
The total current can now be determined from the distribution
function
\begin{equation}
j(t)=-\frac{e}{\tau}\sum_{{\bf k}}{\bar \varepsilon} ({\bf k})
\int_{-\infty}^{t}d{\bar t} e^{-(t-{\bar t})/\tau}f\left ({\bf
k}+e\int_{t}^{{\bar t}}dt^\prime{\bf E}(t^\prime)\right ).
\label{jBoltzmann}
\end{equation}
Shifting the momentum via ${\bf k}\rightarrow {\bf
k}-e\int_{t}^{{\bar t}}dt^\prime{\bf E}(t^\prime)$ and then
integrating over the Brillouin zone gives the following analytical
expression for the current:
\begin{eqnarray}
j(t)&=&-\frac{e}{\sqrt{d}}\frac{eE\tau}{1+e^{2}E^{2}\tau^{2}} \int
d\varepsilon \rho (\varepsilon )\varepsilon f(\varepsilon) \nonumber\\
&\times&\left[ 1-\left(\cos (eEt)-eE\tau\sin (eEt) \right)
e^{-t/\tau} \right] . \label{jBoltzmannsol}
\end{eqnarray}
Analysis of this solution shows  that the current is a strongly
oscillating function of time for $t\ll\tau$ (when $E$ is large)
which then approaches the steady-state value
\begin{equation}
j^{\rm steady}=\frac{eE\tau}{1+e^{2}E^{2}\tau^{2}}j_{0},
\label{jBoltzmannsteady}
\end{equation}
where
\begin{equation}
j_{0}=-\frac{e}{\sqrt{d}}\int d\varepsilon \rho (\varepsilon
)\varepsilon f(\varepsilon ), \label{j0}
\end{equation}
for $t\gg\tau$ (see Fig.~\ref{fig: boltzmann} for an example of the
solutions in the infinite-dimensional limit). It is interesting that
the amplitude of the steady-state current is proportional to $E$ in
the case of a weak field (the linear-response regime), and then
becomes proportional to $1/E$ when the field is strong ($eE\gg
1/\tau$). In this nonlinear regime, the amplitude of the current
goes to zero as the field increases. As it will be shown in the
following Subsection, the exact quantum solution of the problem results in
qualitatively different behavior from the Boltzmann case for the
current as a function of time.

\subsection{Perturbative expansion for the current in the quantum case}

In the quantum case, the electrical current can be found by
calculating the expectation value of the electrical current
operator. This operator is formed from the sum over momentum of the
product of the electric charge times the velocity vector at {\bf k}
times the number operator for electrons in state {\bf k}. Hence, the
time-dependence of the expectation value of the electric current
components is
\begin{equation}
j_{\alpha}(t) =e\sum_{{\bf k}}\frac{\partial}{\partial k_{\alpha}}
\epsilon \left({\bf k}-e{\bf A}(t)\right)
\langle c_{{\bf
k}}^{\dagger}(t)c_{{\bf k}}(t) \rangle , \label{jalpha}
\end{equation}
which becomes
\begin{equation}
j_{\alpha}(t) =-i\frac{e}{\sqrt{d}} \sum_{{\bf k}} \sin \left(
k_{\alpha}-eA_{\alpha}(t)\right) G_{{\bf k}}^{<}(t,t),
\label{jalpha2}
\end{equation}
after using the definition of the lesser Green function.
This Green function can be found
from Eq.~(\ref{GlGK}), where the expressions for the retarded, advanced
and Keldysh Green functions on the r.h.s. of Eq.~(\ref{GlGK})
are given by Eqs.~(\ref{DysonGR2})-(\ref{DysonGK2}).
Since the electric field lies along the lattice diagonal,
all its components are equal to each other,
and the magnitude of the electric current becomes:
\begin{equation}
j(t)=\sqrt{d}j_{\alpha}(t),
\label{j}
\end{equation}
for any component $\alpha$
The momentum summation in Eq.~(\ref{jalpha2})
can be performed by using the definitions of energy functions
in Eqs.~(\ref{eps}) and (\ref{bareps}) and the expression for
joint density of states in Eq.~(\ref{rho2}):
\begin{eqnarray}
j(t) &=&\frac{ie}{2\sqrt{d}} \int d\epsilon\int d{\bar
\epsilon}\rho_{2}(\epsilon ,{\bar\epsilon}) \nonumber\\
&\times&\left[{\bar
\epsilon}\cos \left( eA_{\alpha}(t)\right) -\epsilon\sin \left(
eA_{\alpha}(t)\right) \right]
\nonumber \\
&~&\times\left[
G_{\epsilon ,{\bar \epsilon}}^{K}(t,t)
-G_{\epsilon ,{\bar \epsilon}}^{R}(t,t)
+G_{\epsilon ,{\bar \epsilon}}^{A}(t,t)
\right] .
\label{j2}
\end{eqnarray}
The sum of the last two terms in Eq.~(\ref{j2})
gives zero, since
\begin{eqnarray}
-G_{\epsilon ,{\bar \epsilon}}^{R}(t_{1},t_{2}) +G_{\epsilon ,{\bar
\epsilon}}^{A}(t_{1},t_{2})&=& i\langle\{c_{{\bf k}}(t_{1}),c_{{\bf
k}}^{\dagger}(t_{2}) \}\rangle \nonumber\\
&=&i\delta (t_{1}-t_{2}) \label{moment}
\end{eqnarray}
is momentum-independent and the velocity is an odd function of momentum.

\begin{figure}[h]
\centering{
\includegraphics[width=5.5cm]{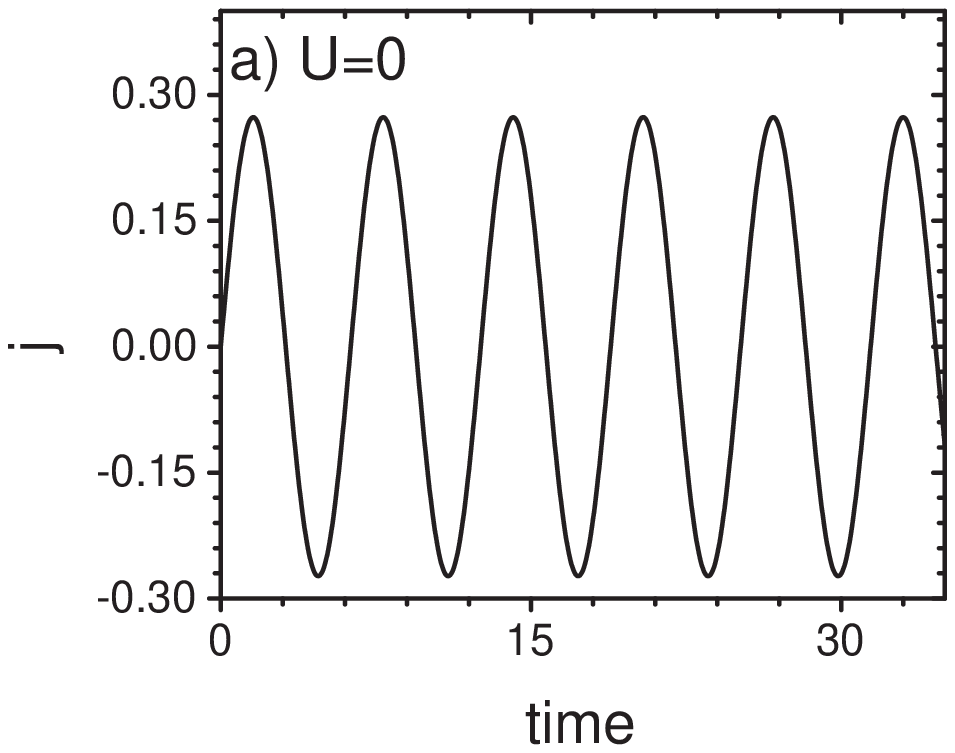}
\includegraphics[width=5.5cm]{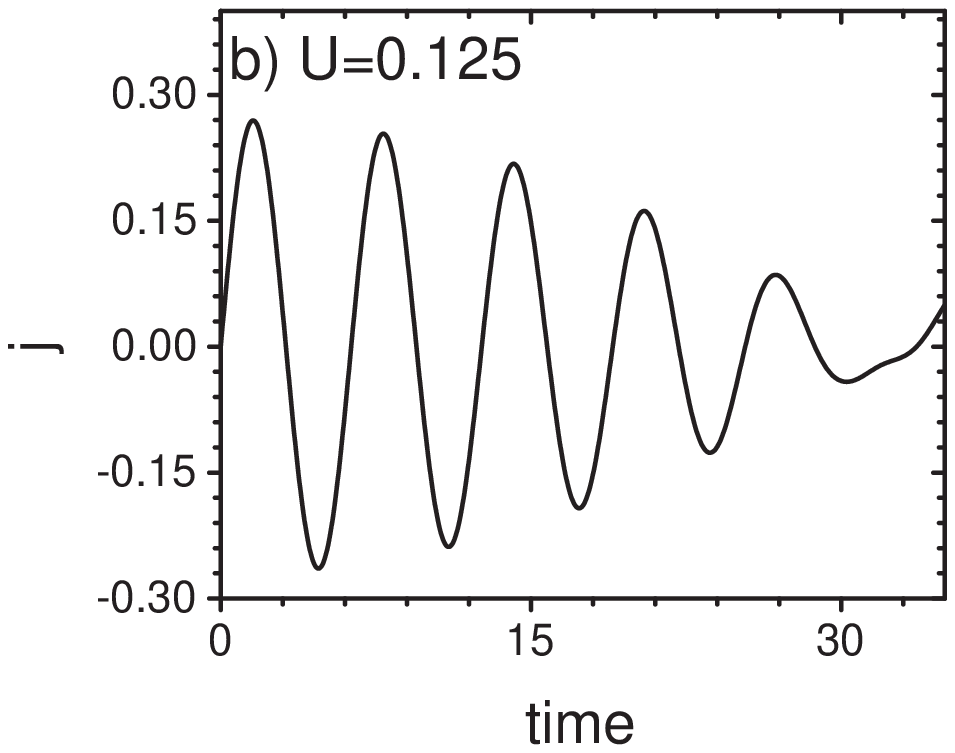}
\includegraphics[width=5.5cm]{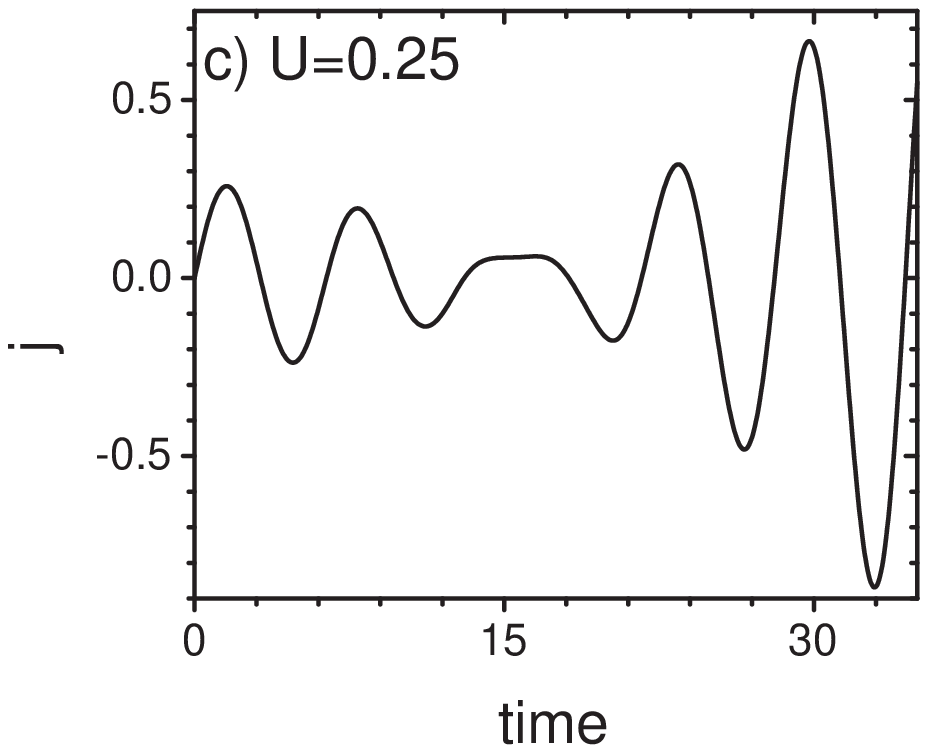}
\includegraphics[width=5.5cm]{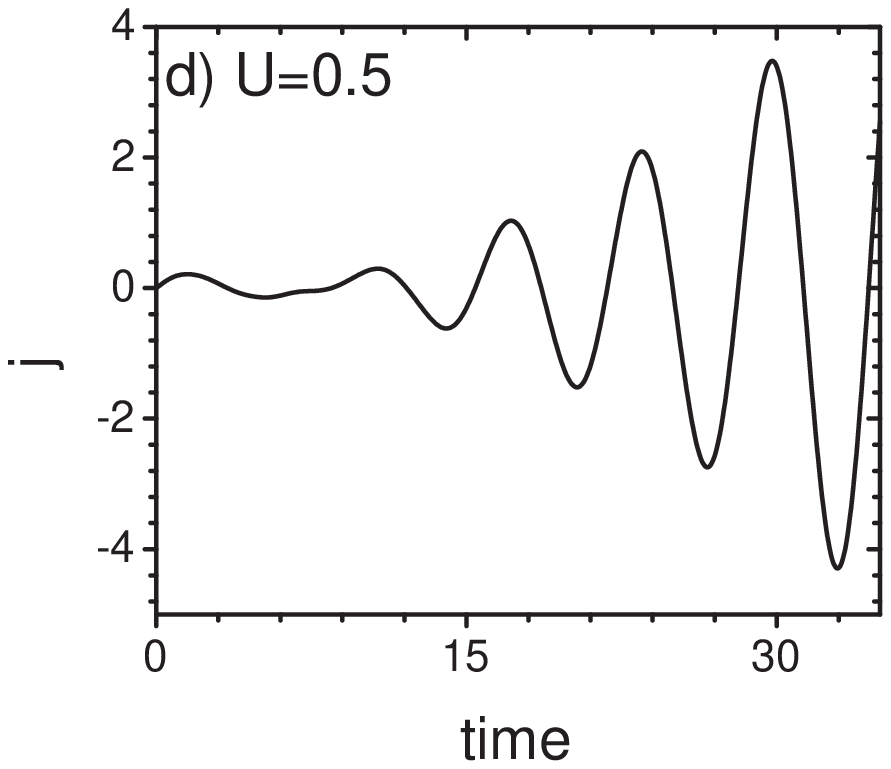}
} \caption{The PT electric current as a function of time for $E=1, \beta
=10$ and different values of $U$. Note the unphysical increase in the current
at long times.} \label{fig:4}
\end{figure}

Therefore,
\begin{eqnarray}
j(t)&=&\frac{ie}{2\sqrt{d}} \int d\epsilon\int d{\bar
\epsilon}\rho_{2}(\epsilon ,{\bar\epsilon}) \left[{\bar
\epsilon}\cos \left( eA_{\alpha}(t)\right) \right .\nonumber\\
&-&\left .\epsilon\sin \left(
eA_{\alpha}(t)\right) \right]\left[ G_{\epsilon ,{\bar
\epsilon}}^{K0}(t,t)\right.
\nonumber \\
&+& \frac{U^{2}}{4}
\int_{-\infty}^{+\infty}dt_{1}\int_{-\infty}^{+\infty}dt_{2}\nonumber\\
&\times& (G_{\epsilon ,{\bar \epsilon}}^{R0}(t,t_{1})
G_{loc}^{R0}(t_{1},t_{2}) G_{\epsilon ,{\bar
\epsilon}}^{K0}(t_{2},t)\nonumber\\
& +& \left .G_{\epsilon ,{\bar
\epsilon}}^{K0}(t,t_{1}) G_{loc}^{A0}(t_{1},t_{2}) G_{\epsilon ,{\bar
\epsilon}}^{A0}(t_{2},t) \right.
\nonumber \\
&+&\left.
G_{\epsilon ,{\bar \epsilon}}^{R0}(t,t_{1})
G_{loc}^{K0}(t_{1},t_{2})
G_{\epsilon ,{\bar \epsilon}}^{A0}(t_{2},t))
\right] ,\nonumber \\
\label{j2a}
\end{eqnarray}
where we restricted to the half-filling case, used
Eq.~(\ref{SigmaPT2}), and neglected the term that is fourth order in
$U$. Integration over ${\bar \varepsilon}$ in Eq.~(\ref{j2}) yields
\begin{eqnarray}
j(t)&=&\frac{e}{\sqrt{d}}\int d\varepsilon \rho (\varepsilon
)\varepsilon f(\varepsilon ) \sin (eA(t))
\nonumber \\
&+&\frac{eU^{2}}{4\sqrt{d}} \int_{-\infty}^{t}
dt_{1}\int_{-\infty}^{t_{1}} dt_{2}
\left[-\frac{i}{2}S(t_{2},t_{1})\cos (eA(t))\right.\nonumber\\
&-& \varepsilon\sin
(eA(t)) \Big]
\exp \left[
-\frac{1}{4}\left\{
C^{2}(t_{2},t_{1})+2S^{2}(t_{2},t_{1})
\right\}
\right]
\nonumber\\
&\times&
\int d\varepsilon \rho (\varepsilon )[2f(\varepsilon )-1]
\exp [-i\varepsilon C(t_{2},t_{1})]
\nonumber \\
&+&\frac{ieU^{2}}{16\sqrt{d}} \int_{-\infty}^{t}
dt_{1}\int_{-\infty}^{t}
dt_{2} \left[S(t_{2},t_{1})\cos (eA(t))\right.\nonumber\\
&-&C(t_{2},t_{1})\sin (eA(t))
\Big]\nonumber\\
&\times&
\exp \left[
-\frac{1}{4}\left\{
C^{2}(t_{2},t_{1})+2S^{2}(t_{2},t_{1})
\right\}
\right]\nonumber\\
&\times& \int d\varepsilon \rho (\varepsilon )[2f(\varepsilon )-1]
\exp [i\varepsilon C(t_{2},t_{1})]
\label{j3}
\end{eqnarray}
In order to simplify the expression, we have introduced the functions
\begin{eqnarray}
C(t_{2},t_{1})=\int_{t_{1}}^{t_{2}}d{\bar t}\cos (eA({\bar t}))
\nonumber \\
S(t_{2},t_{1})=\int_{t_{1}}^{t_{2}}d{\bar t}\sin (eA({\bar t})).
\label{CS}
\end{eqnarray}
In our numerical work, we
consider the current in the case of
a constant electric field $E$ turned on at time $t=0$, i.e.
$A(t)=-Et\theta (t)$. In this case, we find:
\begin{eqnarray}
C(t_{2},t_{1})&=&\theta (-t_{1})\theta (-t_{2})\left[
t_{2}-t_{1}\right] \nonumber\\
&+&\theta (-t_{1})\theta (t_{2}) \left[\frac{\sin
(eEt_{2})}{eE}-t_{1}\right]
\nonumber \\
&+&\theta (t_{1})\theta (-t_{2}) \left[t_{2}-\frac{\sin
(eEt_{1})}{eE}\right] \nonumber\\
&+&\theta (t_{1})\theta (t_{2})
\frac{1}{eE}\left[\sin (eEt_{2})-\sin (eEt_{1})\right] ,
\label{C2} \\
S(t_{2},t_{1})&=&-\theta (-t_{1})\theta (t_{2})
\frac{1}{eE}\left[1-\cos (eEt_{2})\right] \nonumber\\
&-&\theta (t_{1})\theta
(-t_{2}) \frac{1}{eE}\left[\cos (eEt_{1})-1\right]
\nonumber \\
&-&\theta (t_{1})\theta (t_{2}) \frac{1}{eE}\left[\cos
(eEt_{1})-\cos (eEt_{2})\right] . \label{S2}
\end{eqnarray}

\begin{figure}[h]
\centering{
\includegraphics[width=7.0cm]{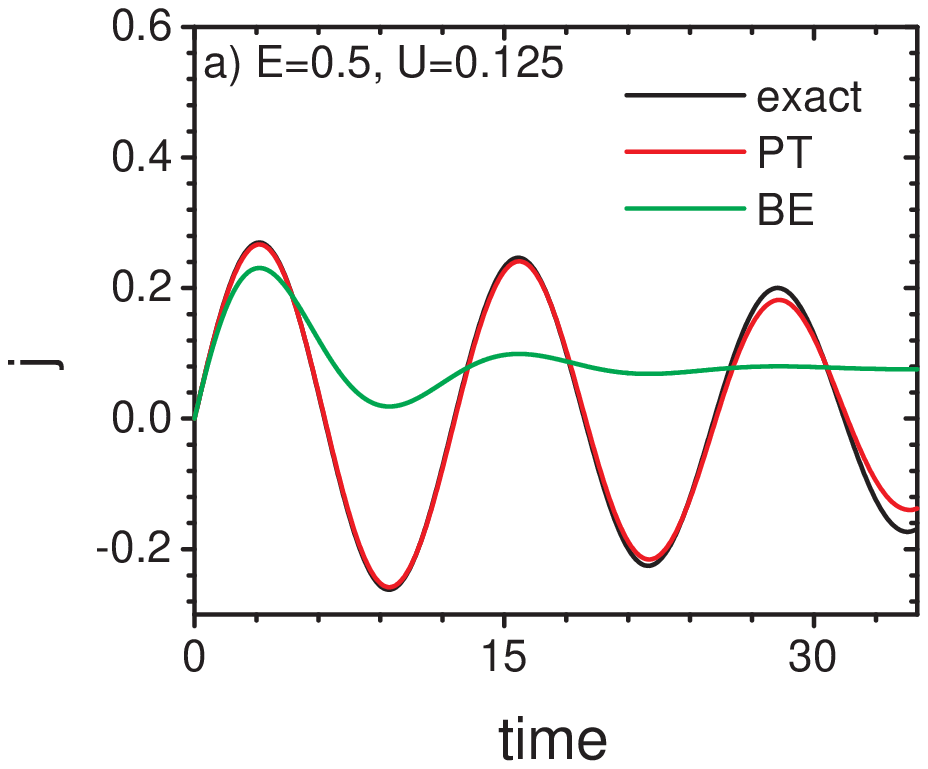}
\includegraphics[width=7.0cm]{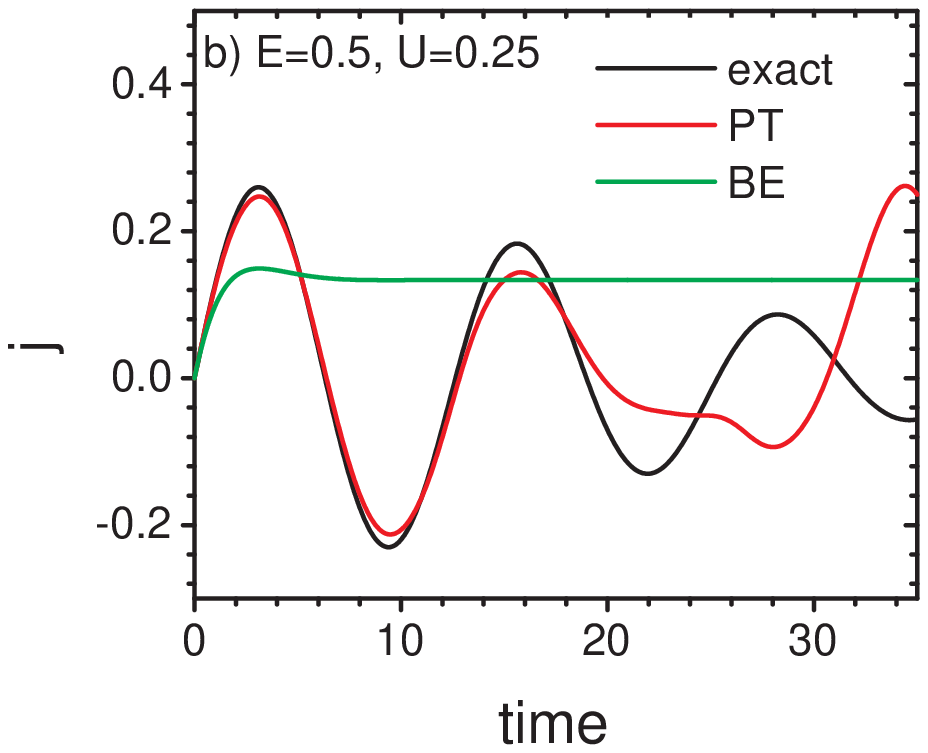}
\includegraphics[width=7.0cm]{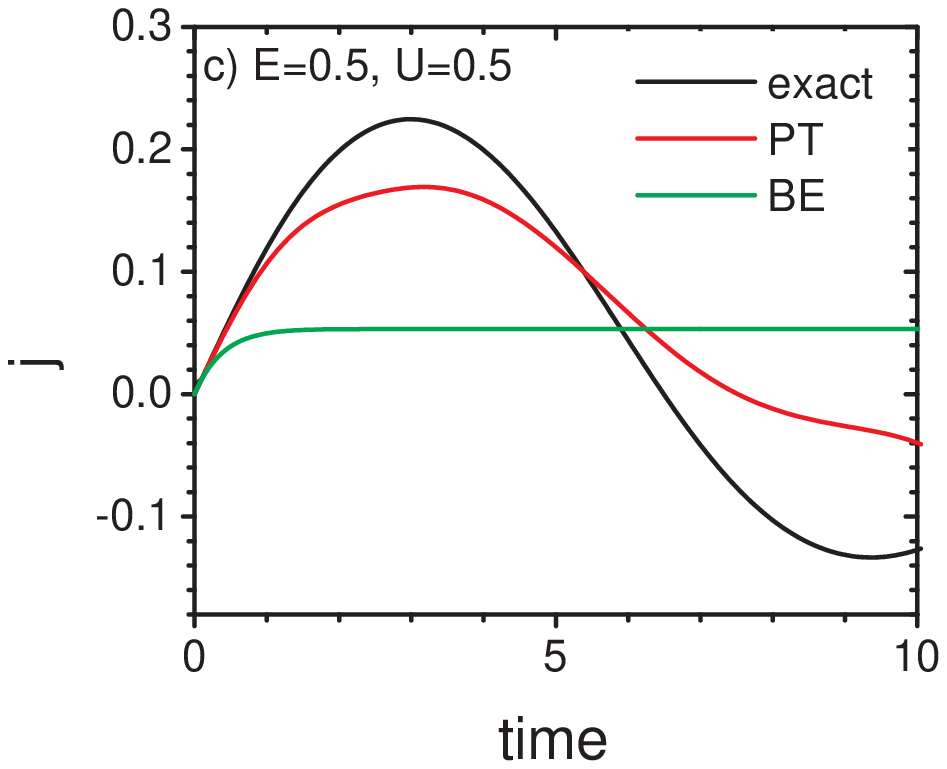}
} \caption{The PT electric current as a function of time for $E=0.5,
\beta =10$ and different values of $U$ (red lines). The exact DMFT
calculation results and the Boltzmann equation solution are
presented by black and green lines, respectively. } \label{fig:5}
\end{figure}

Equations (\ref{j3}), (\ref{C2}) and (\ref{S2}) determine the time
dependence of the current when a constant electric field is turned
on at time $t=0$. It is difficult to find an analytic expression
for $j(t)$, except for the simplest case of $U=0$. In this case:
\begin{eqnarray}
j(t)=j_{0}\sin \left( eEt\right) \label{jU0}
\end{eqnarray}
(the so-called Bloch oscillation\cite{Bloch}), where the amplitude of the
oscillations $j_{0}$ is given by Eq.~(\ref{j0}) and the Bloch
frequency is\cite{Turkowski} $\omega_{B}=eE$. In the limit of
infinite dimensions, the current amplitude $j_{0}$ can also be
written as:
\begin{equation}
j_{0}= \frac{1}{2}\frac{e}{\sqrt{d}}\int d\epsilon \rho (\epsilon )
\frac{d f(\epsilon)}{d\varepsilon} , \label{JU0amplitude}
\end{equation}
after integrating by parts.

\begin{figure}[h]
\centering{
\includegraphics[width=7.0cm]{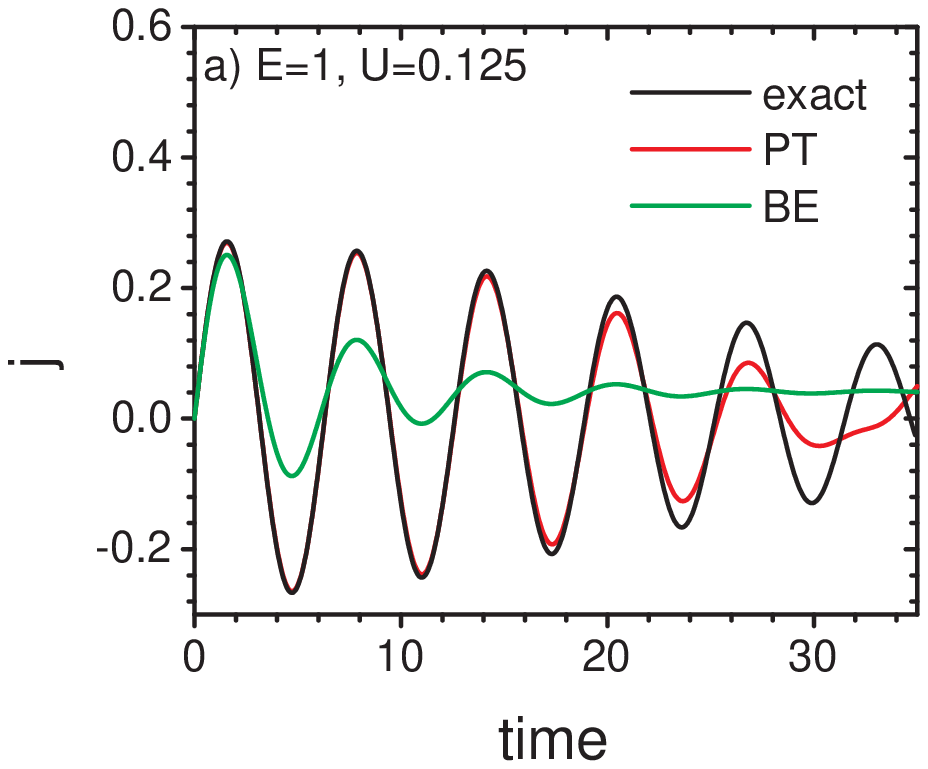}
\includegraphics[width=7.0cm]{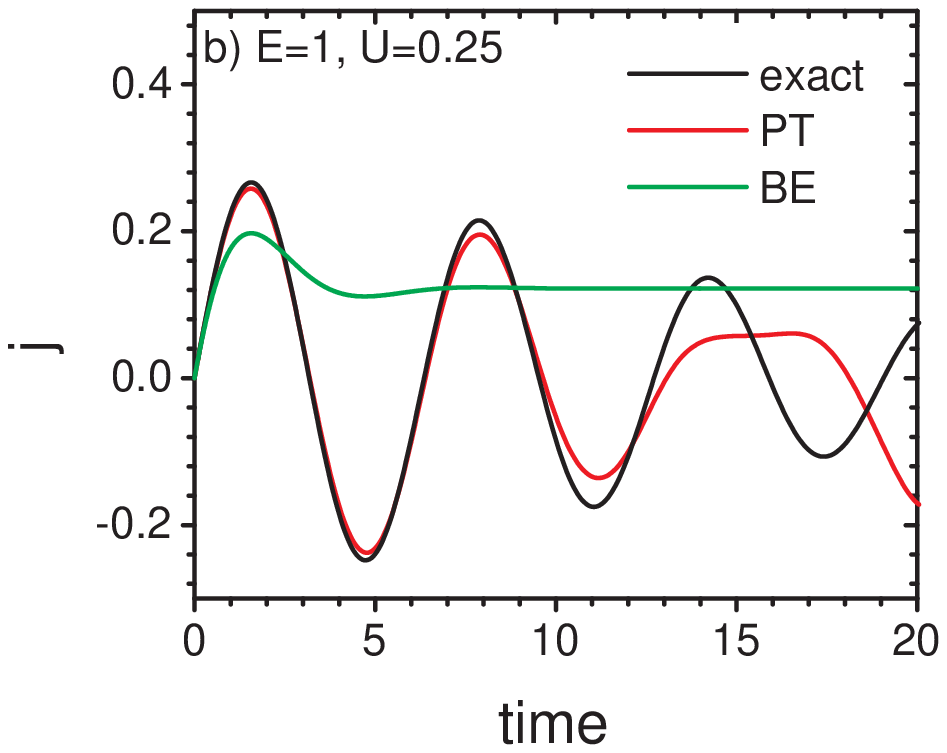}
\includegraphics[width=7.0cm]{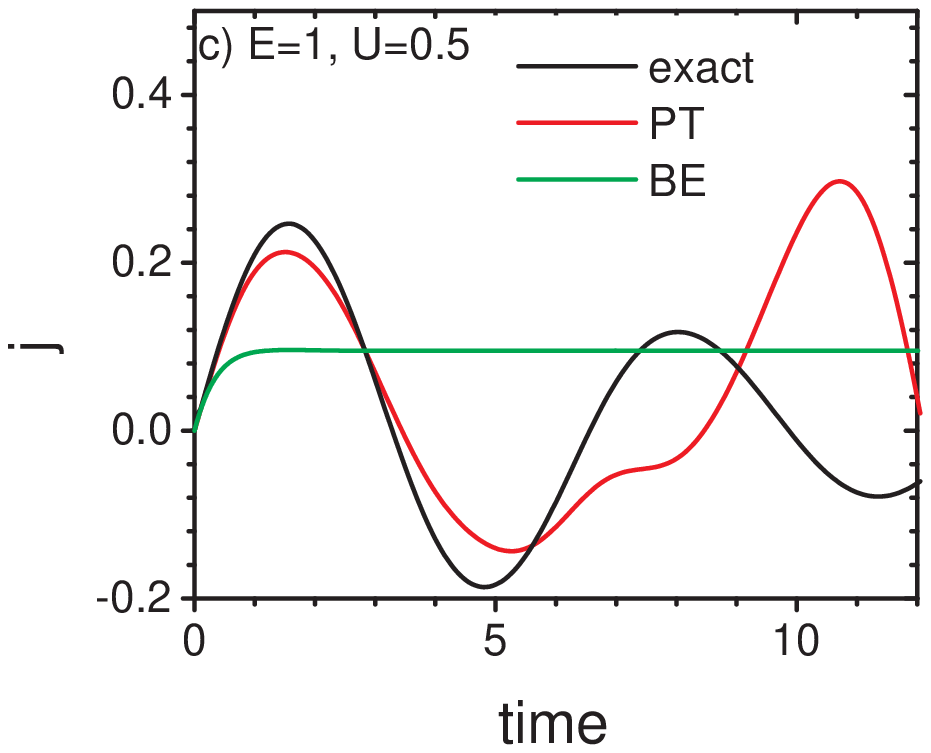}
} \caption{The PT electric current as a function of time for $E=1.0,
\beta =10$ and different values of $U$ (red lines). The exact DMFT
results and the Boltzmann equation solution are
presented by black
 and green lines, respectively.} \label{fig:6}
\end{figure}

Comparison of this result with the general
expression for the current given in Eq.~(\ref{j3})
yields the following formal expression for the current in a strictly truncated
perturbation theory expansion:
\begin{eqnarray}
j(t)=j_{0}\sin \left( eEt\right)+U^{2}j_{2}(t). \label{jU}
\end{eqnarray}
Thus, the electric current is a superposition
of an oscillating part and some other part, whose time-dependence
will be determined below.

\begin{figure}[h]
\centering{
\includegraphics[width=7.0cm]{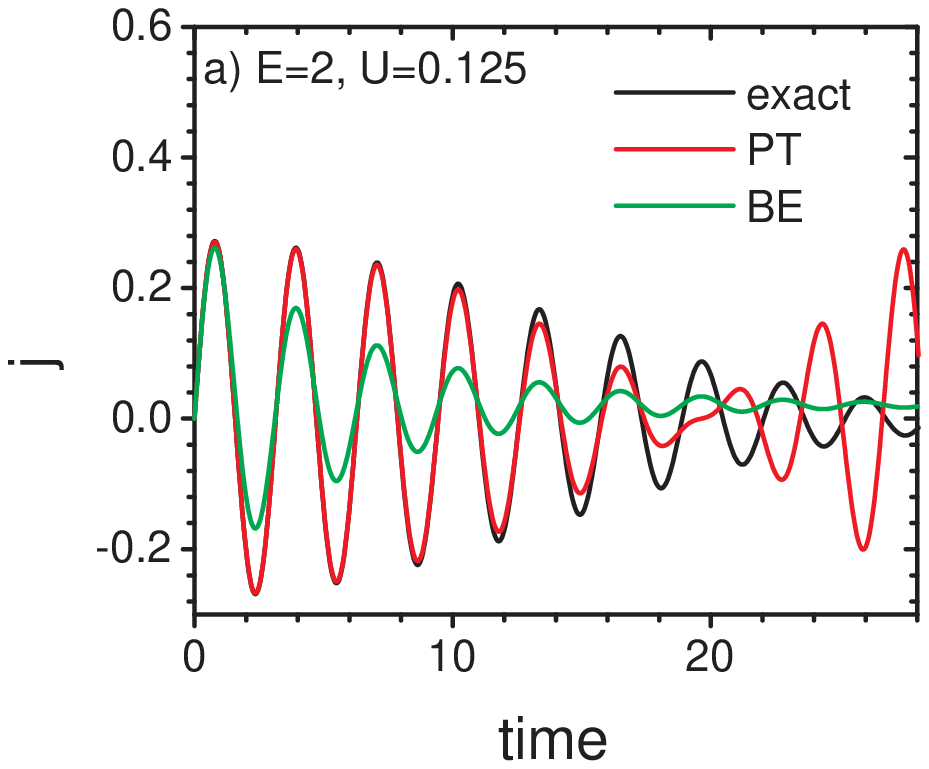}
\includegraphics[width=7.0cm]{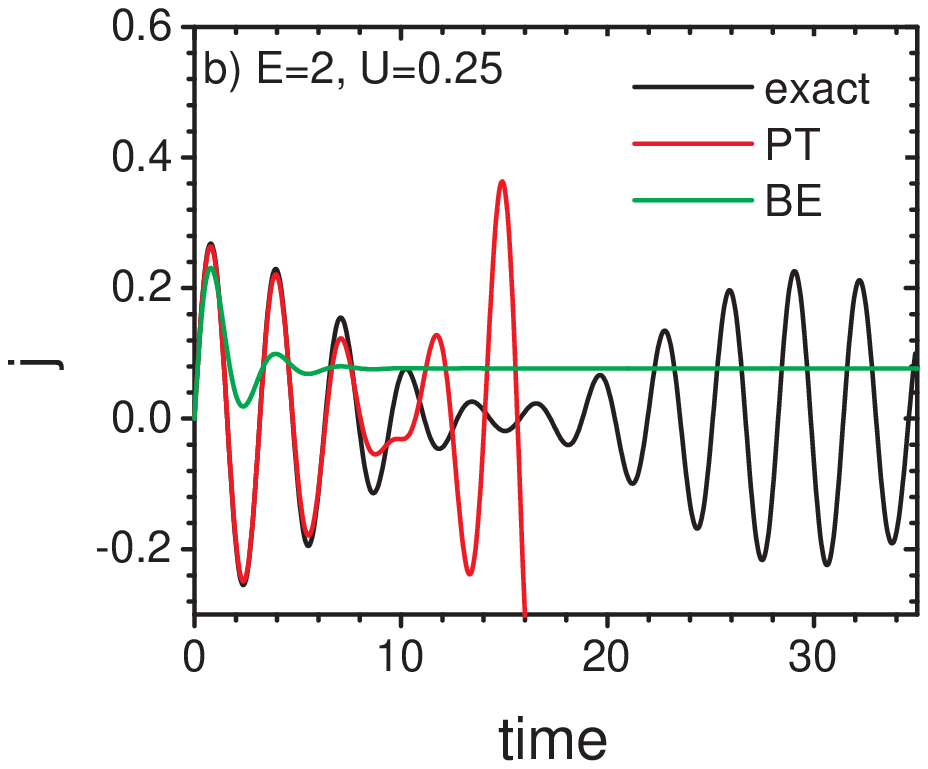}
\includegraphics[width=7.0cm]{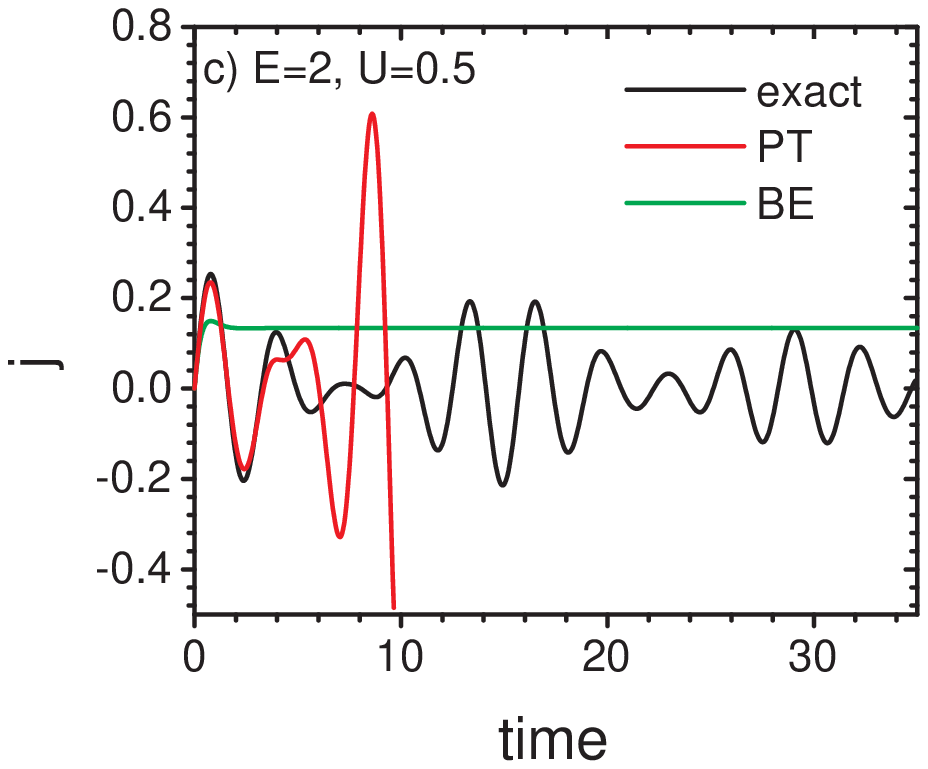}
} \caption{The PT electric current as a function of time for $E=2.0,
\beta =10$ and different values of $U$ (red lines). The exact DMFT
calculation results and the Boltzmann equation solution are
presented by black
 and green lines, respectively. Note that the perturbative values
for the current become so large that they are not shown for larger times.
} \label{fig:7}
\end{figure}

Numerical results for the time-dependence of the electric current
calculated from Eq.~(\ref{j3}) are presented in
Figs.~\ref{fig:4}--\ref{fig:7}. As follows from these figures, the
current oscillates for all time. This is the main difference from
the Boltzmann equation result (Fig.~\ref{fig: boltzmann}), where the
current rapidly reaches a constant steady-state value. As it will be
shown below, the second order perturbation theory cannot describe
the steady-state regime correctly. However, the perturbation theory
should give reasonable results for times smaller than $\sim 2/U$,
which we consider below. As depicted in Figure~\ref{fig:4}, the current
depends only weakly on $U$, except for the case $U\sim 0.5$. This is
because the current is dominated by the Bloch term $j_{0}$ [see
Eq.~(\ref{jU})]. As time increases, the amplitude of the
oscillations decrease. This decrease is proportional to $U^{2}$, in
agreement with Eq.~(\ref{j3}). As $t$ increases further, the current
amplitude goes through zero (at a time we call $t_{\rm pert}$), and
then starts to increase dramatically. This is an unphysical result,
so we assume that the perturbation theory is not valid for times
larger than $t_{\rm pert}$, i.e. the time when the perturbation
theory breaks down. This time is always smaller than the time needed
to see the steady state develop, so this perturbation theory is not
capable of producing the steady-state response.  The period of the
oscillations remains essentially equal to the Bloch oscillation
period, but it is not well defined for damped oscillations.

The validity of the second-order nonequilibrium perturbation theory
can also be checked by comparing the results for the current with
exact numerical results calculated by the nonequilibrium DMFT method
\cite{Nashville,Turkowski1,dmft_fk}. We present the corresponding
results for cases of different values of $U$ and $E$ in
Figs.~\ref{fig:5}--\ref{fig:7}. As shown in these Figures, the
perturbation theory results are close to the exact results only for
times smaller than $\sim 2/U$. Note that although the perturbation
theory does show an increase in the current for intermediate times,
it does not properly show the quantum beats in the current, which
occur, with a beat period on the order of $1/U$, even though the
shape of the curve is qualitatively similar.  These beats occur only
for large fields ($E\ge 2$ here).

In Figures~\ref{fig:5}--\ref{fig:7}, we also present the results for
the current in the case of the Boltzmann equation solution
Eq.~(\ref{jBoltzmannsol}), where the scattering time is fixed at the
quantum Boltzmann equation value
$\tau =\rho (\mu )/(4\pi |{\rm Im}\Sigma (\omega =0)|)$
(see, for example Ref.~\onlinecite{Kiel}; this result is the transport
relaxation time). Substitution of
Eq.~(\ref{ImSigmaequilibrium}) into this expression yields:
\begin{equation}
\tau =1/(\pi^{2}U^{2}).
\label{tauestimate}
\end{equation}
The Boltzmann equation approach shows a rapid approach to the
steady state, much faster than what is seen in the exact numerical
results or in the perturbation theory.  Although the Boltzmann equation
is accurate at small times, it is clear that the quantum system has much
richer behavior than what is predicted by the semiclassical approach,
at least in cases where the electric field is large. Note that the
functional form of the semiclassical Boltzmann equation can fit the exact
results for the current much better if we adjust the relaxation time
$\tau$ to yield the best fit of the data rather than use Eq.~(\ref{tauestimate}),
but the relaxation time then becomes just a fitting parameter and is not derived
from a microscopic model.

To complete the analysis of the time-dependence of the electric
current, we present an analytical expression for the electric
current in the limit of strong electric fields $E\rightarrow\infty$.
Substitution of Eqs.~(\ref{C2}) and (\ref{S2}) into Eq.~(\ref{j3})
gives the following result for the electric current in the limit of
large electric fields:
\begin{eqnarray}
j(t)\simeq &-&\frac{e}{\sqrt{d}}\int d\varepsilon \rho (\varepsilon
) \varepsilon
f(\varepsilon ) \nonumber \\
&\times& \left[1-U^{2}B(\beta )-\frac{U^{2}}{4}t^{2} \right] \sin
(eEt),
\nonumber \\
\label{jstrongfields}
\end{eqnarray}
where
\begin{eqnarray}
B(\beta )= \left[ \int d\varepsilon \rho (\varepsilon )\varepsilon
f(\varepsilon )\right]^{-1} \int_{0}^{\infty}dx
\int_{x}^{\infty}dyye^{-y^{2}} \nonumber \\
\times \int d\varepsilon \rho (\varepsilon ) f(\varepsilon )
\sin(2\varepsilon y). \label{B}
\end{eqnarray}

Numerical calculations show that this is a positive decreasing
function of temperature: $0.25<B(\beta )<0.5$. As follows from
Eq.~(\ref{jstrongfields}), the electron-electron correlations lead
to a decrease of the current amplitude at short times, since the
terms proportional to $U^{2}$ in the square brackets in
Eq.~(\ref{jstrongfields}) have a minus sign in front of it. This
correction is not large at short times, since we consider the case
$U\ll 1$. However, from the numerical calculations, we find that
the term in Eq.~(\ref{jstrongfields}) which is proportional to
$U^{2}t^{2}/4$, is dominant in the case of longer times. In fact,
the current is an oscillating function of time. The oscillation
amplitude, decreases with time as $1-U^{2}B(\beta )-U^{2}t^{2}/4$.
At the time $t_{\rm pert}=(2/U)\sqrt{1-U^{2}B(\beta )} \simeq 2/U$ the
sign in front of the amplitude changes, and the current oscillates
out of phase with the $U=0$ case. It is important that the period of
the oscillations is equal to the period of the Bloch oscillations in
the noninteracting case. At longer times, the term proportional to
$t^{2}$ is the most important and the time dependence is
$j(t) \sim t^{2}\sin (eEt)$ at times longer than $\sim2/U$. These
results are in a qualitative agreement with the numerical results
for the perturbation theory presented in Figs.~\ref{fig:4}--\ref{fig:7}.
However, it is plain to see that the a growing amplitude for the current
as time increases for $t>2/U$ is an unphysical result.

Unfortunately, it is impossible to find an analytical expression for
the current in the limits of intermediate and small
fields. However, our numerical analysis shows that the current
qualitatively has the dependence given
in Eq.~(\ref{jstrongfields}).  In particular, the steady state is never
reached before the perturbation theory develops pathological behavior.

\section{Local density of states}

The time-dependent density of states can be calculated from the local
retarded
Green function in the following way:
\begin{equation}
A(\omega ,T)=-\frac{1}{\pi}{\rm Im}G^{R}(\omega ,T), \label{Aw}
\end{equation}
where $G^{R}(\omega ,T)$ is the Fourier transform of the retarded
Green function $G^{R}(t_{1},t_{2})$ with respect to the relative
time coordinate $\tau =t_{1}-t_{2}$, and $T$ is the average time
coordinate $T=(t_{1}+t_{2})/2$. The perturbation theory retarded
Green function is determined in Eq.~(\ref{DysonGR2}). Integration
over the Brillouin zone and substitution into Eq.~(\ref{Aw}) yields:
\begin{eqnarray}
A(\omega ,T)&=&\frac{1}{\pi}
\int_{0}^{\infty}d\tau \cos (\omega\tau )\nonumber\\
&\times& \exp\left[ -\frac{1}{4}
\left(
C^{2}(T+\tau /2,T-\tau /2)\right.\right.\nonumber\\
&+&\left.S^{2}(T+\tau /,T-\tau /2) \right)\Bigr]
\nonumber \\
&\times& \Bigr\{ 1-\frac{U^{2}}{4} \int_{T-\tau /2}^{T+\tau
/2}dt_{3}\int_{T-\tau /2}^{t_{3}}dt_{4}
\nonumber\\
&\times&\exp\left[-\frac{1}{2} \left(
C^{2}(t_{3},t_{4})+S^{2}(t_{3},t_{4}) \right) \right]
\nonumber \\
&\times& \exp\left[\frac{1}{2} \left( C(T+\tau /2,T-\tau
/2)C(t_{3},t_{4})\right. \right.\nonumber\\
&+&\left. S(T+\tau /2,T-\tau /2)S(t_{3},t_{4}) \right) \Bigr]
\Bigr\} , \label{Aw2}
\end{eqnarray}
where $C(t_{3},t_{4})$ and $S(t_{3},t_{4})$ are defined in
Eqs.~(\ref{C2}) and (\ref{S2}).

Numerical calculations of the time dependence of the density of
states show that the PT correction is small for times less than
$t_{\rm pert}$ $\sim 2/U$. Similar to the case\cite{Turkowski} $U=0$, the
density of states develops peaks at the Bloch
frequencies. These peaks grow as time increases, essentially due to
the zeroth-order DOS. However, these peaks do not split into two
peaks as time increases for times less than $\sim 2/U$, except for the
zero frequency peak. This result is different from the exact DMFT
calculations\cite{Turkowski1} at $U=0.5$.
Generically, one expects the DOS to broaden as scattering is added, and to split,
by an energy of the order of $U$ in the steady state due to the interactions;
the numerical calculations indicate that this is indeed what actually
happens even though the truncated perturbation theory does not explicitly show
it.

It is interesting that one can also find
analytical expressions for the zeroth and the first two spectral
moments, which are time independent (the first and second-order moments
are written for the half-filling case):
\begin{eqnarray}
\int d\omega A(\omega ,t)=1,
\nonumber \\
\int d\omega \omega A(\omega ,t)=0,
\nonumber \\
\int d\omega \omega^{2} A(\omega ,t)=\frac{1}{2}+\frac{U^{2}}{4}.
\end{eqnarray}
These results coincide with the exact results obtained in
Ref.~\onlinecite{Turkowski1} at half-filling.

\section{Conclusions}

We have studied the response of the conduction electrons in
the spinless Falicov-Kimball model to an external electric field by
using a strictly truncated perturbative expansion to second order in $U$.
We derived
explicit expressions for the retarded, advanced and Keldysh Green
functions and used them
to calculate the time-dependence of the electric current and the density
of states.  We examined
the case when a constant electric field is turned on at a
particular moment of time. We found  that Bloch oscillations
of the current are present up to times at least as long as
$t\sim 2/U$, but with a decreasing amplitude. In the perturbative approach,
the oscillations are around a zero average current and the PT breaks down
for longer times, so we cannot reach the steady state.
This result is different from that of the semiclassical Boltzmann equation
approach.  There, the Bloch oscillations rapidly decay with
time and the system approaches the steady state with
$j={\rm const.}$ as time goes to infinity. Increasing the Coulomb repulsion
reduces the amplitude of the oscillations more rapidly.
The perturbation theory and the
Boltzmann equation results for the electric current are close
for short times.

The perturbation theory results for the Green functions can be used
to test the accuracy of the numerical nonequilibrium DMFT
calculations \cite{Nashville,Turkowski1,dmft_fk} in the case of weak
correlations, when the system is not in the insulating regime. We also found
excellent agreement for short times, but the PT becomes ill behaved at
long times.

Our results can be generalized to more
complicated cases. In particular, cases with
different time-dependence for the field (pulses, periodic fields etc.)
can be considered.
Also higher order terms can be taken into account in the PT, or one can make the
PT self-consistent. Finally, we can examine other strongly correlated models
as well,
where we expect similar behavior.

The most important result of this work is that it shows how
difficult it is to accurately determine the steady-state response in
a perturbative approach. At the very least, one needs to perform a
self-consistent perturbation theory to be able to properly determine
the long-time behavior, but, since we know that equilibrium PT is
most inaccurate at low frequencies, it is natural to conclude that
nonequilibrium PT is most inaccurate at long times.  Hence, our work
shows explicitly how challenging it is to try to determine the
steady states of quantum systems unless one can formulate a
steady-state theory from the start.  Work along those lines is in
progress for the exact solution via DMFT.

\section*{Acknowledgments}
V.~T. would like to thank Natan Andrei for a valuable discussion. We
would like to acknowledge support by the National Science Foundation
under grant number DMR-0210717 and by the Office of Naval Research
under grant number N00014-05-1-0078. Supercomputer time was provided
by the DOD HPCMO at the ASC and ERDC centers. We would like to thank
the Referee of our paper\cite{Kiel}, which appeared in the
proceedings of the Workshop on Progress in Nonequilibrium Physics
III (Kiel, Germany, August, 2005), for suggesting that we examine
the Boltzmann equation solution for the current and compare it to
solutions found with other methods.

\appendix

\section{Expansion for the time-ordered equilibrium self-energy to second order
in U}

The time-ordered self-energy $\Sigma_{lm}(t_{1},t_{2})$
can be expanded by examining an order-by-order solution of the Dyson equation
(we suppress the $T$ superscript in these formulas)
\begin{eqnarray}
&~&G_{ij}(t,t')=G_{ij}^{0}(t,t')
\nonumber\\
&+&\sum_{lm}\int dt_{1}\int dt_{2}
G_{il}^{0}(t,t_{1})\Sigma_{lm}(t_{1},t_{2})G_{mj}(t_{2},t'),\\
&=&G_{ij}^{0}(t,t')
\nonumber\\
&+&\sum_{lm}\int dt_{1}\int dt_{2}
G_{il}^{0}(t,t_{1})\Sigma_{lm}^{(1)}(t_{1},t_{2})G^0_{mj}(t_{2},t')\nonumber\\
&+&\sum_{lm}\int dt_{1}\int dt_{2}
G_{il}^{0}(t,t_{1})\Sigma_{lm}^{(2)}(t_{1},t_{2})G_{mj}^0(t_{2},t')\nonumber\\
&+&\sum_{lmnr}\int dt_{1}\int dt_{2}\int dt_3 \int dt_4
G_{il}^{0}(t,t_{1})\nonumber\\
&\times&\Sigma_{lm}^{(1)}(t_{1},t_{2})
G_{mn}^0(t_2,t_3)\Sigma^{(1)}_{nr}(t_3,t_4)G_{rj}^0(t_{4},t'),
\label{Sigmadef}
\end{eqnarray}
where the second equation explicitly shows the systematic expansion
to order $U^2$ in terms of the first-order $\Sigma^{(1)}$ and
second-order $\Sigma^{(2)}$ perturbative self-energies. The
self-energies are found by a straightforward perturbative expansion
of the evolution operator, when expressed in the interaction
picture:
\begin{eqnarray}
iG_{ij}(t,t')&=&\sum_{\nu =0}^{\infty} \frac{(-i)^{\nu}}{\nu !} \int
dt_{1}...\int dt_{\nu} \nonumber\\
&\times&\langle {\hat T}[{\hat H}_{U}(t_{1})...{\hat
H}_{U}(t_{\nu}) c_{i}(t)c_{j}^{\dagger}(t')]  \rangle_{\rm con},
\label{expansion}
\end{eqnarray}
where the interacting part of the Hamiltonian is
\begin{eqnarray}
{\hat H}_{U}(t_{1})&=&
-[\mu-\mu^{(0)}]\sum_{i}c_{i}^{\dagger}c_{i}
-[\mu_{f}-\mu_f^{(0)}]\sum_{i}f_{i}^{\dagger}f_{i}\nonumber\\
&+&U\sum_{i}f_{i}^{\dagger}f_{i}c_{i}^{\dagger}c_{i}
\label{HU}
\end{eqnarray}
and the statistical average is taken with respect
to the free Hamiltonian
${\hat H}_{0}=-\sum_{\langle ij\rangle}t_{ij}c_{i}^{\dagger}c_{j}^{}
-\mu^{(0)}\sum_ic_{i}^{\dagger}c_i^{}-\mu_f^{(0)}\sum_if_i^\dagger f_i^{}$.
The subscript ``con'' indicates that we take only the connected contractions
generated by Wick's theorem, which must connect the conduction electron
operators for all time values in the matrix elements; the connected
diagrams are included because the disconnected diagrams cancel from a
similar expansion of the partition function, which appears in the
denominator of the matrix-element average.
We have included the terms with the shift of the chemical potentials
into Eq.~(\ref{HU}), since $(\mu-\mu^{(0)})\sim(\mu_{f}-\mu_f^{(0)})\sim U$.
Expanding the electron Green function in Eq.~(\ref{expansion})
up to second order in the interaction and using the definition
of the self-energy in Eq.~(\ref{Sigmadef}), one finds
\begin{equation}
\Sigma^{(1)}_{lm}(t_1,t_2)=(Un_f-\mu+\mu^{(0)})\delta_{lm}\delta(t_1-t_2),
\label{eq: sigma_first}
\end{equation}
and
\begin{equation}
\Sigma^{(2)}_{lm}(t_1,t_2)=U^2G^0_{lm}(t_1-t_2)F^0_{lm}(t_1-t_2)
F^0_{ml}(t_2-t_1),
\label{eq: sigma_second}
\end{equation}
which depends on the difference of the times because we are in equilibrium.
Since the $f$-electron Green function is local, the second-order
self-energy must have $l=m$.  Furthermore, since all of the Green functions
are time-ordered Green functions, the product of the two $f$-electron
Green functions is equal to $n_f(1-n_f)$.

\end{document}